\documentclass[pra,aps,superscriptaddress,footinbib,twocolumn]{revtex4-2}
\usepackage{color}
\usepackage{xcolor}
\usepackage{amsmath}
\usepackage{amsthm}
\usepackage{amssymb}
\usepackage{stmaryrd}
\usepackage{graphicx}
\usepackage{mdframed}
\usepackage{paracol}
\usepackage{siunitx}
\usepackage[unicode=true,
 bookmarks=true,bookmarksnumbered=false,bookmarksopen=false,
 breaklinks=false,pdfborder={0 0 1},backref=false,colorlinks=true]
 {hyperref}

\usepackage[default,authormarkup=none]{changes} 
\usepackage{xcolor} 

\definechangesauthor[color=blue]{refree1}
\definechangesauthor[color=red]{refree2}

\hypersetup{
 linkcolor=magenta, urlcolor=blue, citecolor=blue, pdfstartview={FitH}, hyperfootnotes=false, unicode=true}

\makeatletter

\usepackage{amsfonts}\usepackage{tabularx}\usepackage{dcolumn}\usepackage{bm}\usepackage{graphicx}\usepackage{epstopdf}
\usepackage{times}
\usepackage{cleveref}
\hypersetup{urlcolor=blue}

\makeatother

\newmdtheoremenv{theo}{Theorem}
\begin{document}


\title{Cavity Optomechanical Probe of Gravity Between Massive Mechanical Oscillators}

\author{Ziqian Tang}
\affiliation{Beijing Key Laboratory of Fault-Tolerant Quantum Computing, Beijing Academy of Quantum Information Sciences, Beijing 100193, China}
\affiliation{Bejing National Laboratory for Condensed Matter Physics, Institute of Physics, Chinese Academy of Sciences, Beijing 100190, China}
\affiliation{University of Chinese Academy of Sciences, Beijing 100049, China}

\author{Wenlong Li}
\affiliation{School of mathematics and physics, Qinghai University, Xining 810016, China}
\affiliation{Beijing Key Laboratory of Fault-Tolerant Quantum Computing, Beijing Academy of Quantum Information Sciences, Beijing 100193, China}

\author{Huanying Sun}
\affiliation{Beijing Key Laboratory of Fault-Tolerant Quantum Computing, Beijing Academy of Quantum Information Sciences, Beijing 100193, China}

\author{Xiaoxia Cai}
\affiliation{Institute of High Energy Physics, Chinese Academy of Sciences, Beijing 100049, China}

\author{Tiefu Li}
\affiliation{School of Integrated Circuits and Frontier Science Center for Quantum Information, Tsinghua University, Beijing 100084, China}
\affiliation{Beijing Key Laboratory of Fault-Tolerant Quantum Computing, Beijing Academy of Quantum Information Sciences, Beijing 100193, China}

\author{Yulong Liu}
\email{liuyl@baqis.ac.cn}
\affiliation{Beijing Key Laboratory of Fault-Tolerant Quantum Computing, Beijing Academy of Quantum Information Sciences, Beijing 100193, China}
\date{\today}

\begin{abstract}
Exploring gravitational interactions between objects with small masses has become increasingly timely. Concurrently, oscillators with masses ranging between milligrams and grams in cavity optomechanical systems sparked interest for probing gravity, and even investigating gravity induced decoherence within massive mechanical systems. Here, we present a measurement scheme for probing gravity in a microwave optomechanical setup that incorporates periodic gravitational modulation between the test mass and the driven source mass at the milligram scale. Optomechanically induced transparency (OMIT) can be utilized to sense the gravitational interactions between test masses and source masses. Specifically, the relative variation in the height of the OMIT peak, expressed as $|1 + re^{i\phi}|^2 - 1$, where $r$ represents the ratio of the amplitude of the gravitational driving force to the radiation pressure force of the probe tone, and $\phi$ denotes their phase difference, can reach up to 2.3\% under plausible experimental conditions. This work may facilitate cavity-optomechanical probing of gravitational coupling between milligram-scale mechanical oscillators, a mass regime where modifications to gravity, evidence for extra dimensions, or gravity induced entanglement may emerge.
\end{abstract}

\maketitle


\section{\label{sec:level1}Introduction}
Gravity, as one of the fundamental forces in nature, has always been a focal point for physicists. Starting from Cavendish's torsion balance experiment~\cite{cav1798} in 1798, continuous efforts have been made to explore methods for measuring gravity between objects of different masses to validate the applicability of Newton's law of gravity at different mass scales. In the past two centuries, various experimental schemes based on the interaction between test mass and source mass have been developed, including the incorporation of different geometries such as spheres~\cite{cav1798, Reich1838, Poynting1891, Boys1895, Burgess1899, Zahradnicek1933, Dousse1987, Brack2023}, cylinders~\cite{Wilsing1889, Braun1897, Renner1970, Michaelis1995, Yang1991, Nolting1999, Quinn2013, Lamporesi2008, Luo1998, Hu2005, Schwarz1999, Luo2009, Tu2010, Parks2010, Baldi2005, Fixler2007, Schlamminger2006}, rectangular blocks~\cite{Jolly1881, Richarz1898} etc., as well as different materials such as water~\cite{Yang1991, Nolting1999}, mercury~\cite{Schlamminger2006}, lead~\cite{Jolly1881, Richarz1898, Roland1896},  tungsten~\cite{Rose1969, Luther1982}, stainless steel~\cite{Sagitov1979, Li2018}, brass~\cite{Speake1983, Liu1987, Quinn2001, Kleinvoss2002, Rosi2014}, etc. However, due to the extremely weak nature of gravity, the masses involved in these experimental schemes have mainly been in the kilogram scale for a considerable period. In 1990, Ritter et al. successfully measured the gravity between two Dy-Fe cylinders, each with a mass of about 90 gram, in an experiment testing torsion in general relativity~\cite{Ritter1990}. Furthermore, Tan et al. tested the gravitational inverse-square law at the submillimeter range using two approximately 1-gram I-shaped pendulums with dual modulation and compensation~\cite{PhysRevLett.116.131101Tan,PhysRevLett.124.051301Tan}. The possible deviations from the inverse-square law have not been discovered in the above mass regimes, and merely increasing the mass scale does not guarantee greater discovery potential. Instead, the hope of uncovering modifications of gravity due to small extra dimensions or evidence of new forces lies in progressively smaller mass ranges, with research interest shifting towards measuring the gravitational interactions between milligram-scale oscillators~\cite{Sofia2025}. Subsequently in 2021, Westphal et al., successfully measured the gravity of a single-source mass of 92 milligram, marking significant progress in the field of gravity measurement for small mass objects~\cite{Westphal2021}.
    
\begin{figure*}
    \centering
    \includegraphics[width=0.9\textwidth]{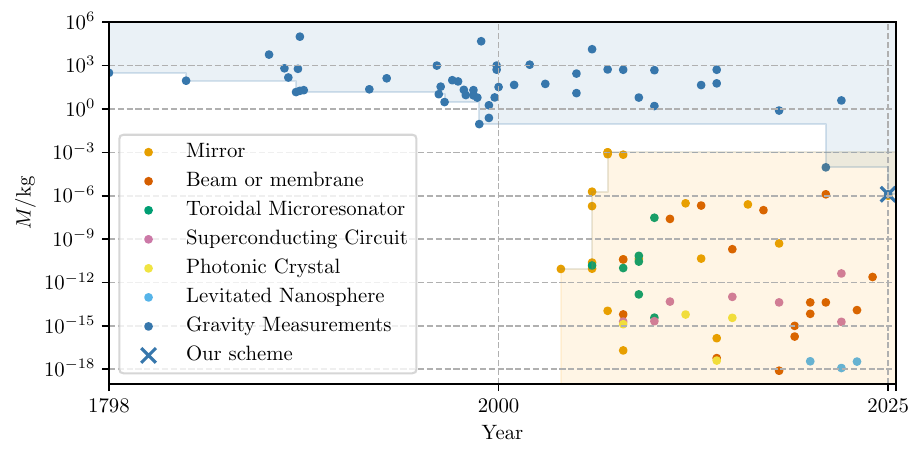} 

    \caption{\label{fig:fig1}The values of source masses (blue marked area) used in gravitational constant measurement experiments and the masses of mechanical resonators (yellow marked area) used in cavity optomechanical experiments from 1790 to 2025 according to Table~\ref{tab:tableG} and Table~\ref{tab:tableOPTO} in Appendix~\ref{sec:tablesGandOPTO}. The horizontal axis represents the year, and the vertical axis represents the mass. It can be seen from the figure that the mass scales involved in both types of experiments overlapped in 2021. The mass scale in our scheme is presented in the figure by a blue ``$\times$'' marker.}
\end{figure*}

Simultaneously, the rapid development of cavity optomechanics has provided new possibilities for probing gravitational forces between increasingly smaller masses. Optomechanics not only allows the observation of macroscopic quantum phenomena, such as the non-classical motion of mechanical resonators~\cite{Korppi2018,Riedinger2018}, but also demonstrates excellent capabilities in measuring small forces~\cite{Fogliano2021} and small displacements~\cite{Liu2020}. Due to these characteristics, it is natural to expect that it will serve as an appropriate platform to measure weak gravitational forces between small objects and even to study gravity within macroscopic quantum systems in the future~\cite{Marletto2017, Miao2020, Westphal2021, Liu2021, Sofia2025, Bose2025, Tang2025}. During the past 20 years, oscillators of different effective masses have been achieved in various cavity optomechanical systems, such as mirrors~\cite{Metzger2004,Arcizet2006,Gigan2006,Corbitt2007,Sofia2025}, beams or membranes~\cite{Kleckner2006,Thompson2008,Kuhn2011,Torres2013}, toroidal microresonators~\cite{Schliesser2006,Anetsberger2010,Park2009}, superconducting circuits~\cite{Regal2008,Rocheleau2010,Teufel2011}, photonic crystals~\cite{Li2008,Zheng2012,Paraiso2015}, and levitated nanospheres~\cite{Delić2020,Militaru2022,Piotrowski2023}, etc. 
As shown in Fig.~\ref{fig:fig1}, there is a noticeable trend of decreasing source masses in gravitational detection experiments, while the masses of macroscopic mechanical devices in cavity optomechanical systems undergo an increase. This convergence has allowed the mass scales used in gravity detection experiments~\cite{Westphal2021} to overlap with those of macroscopic mechanical devices in optomechanical systems~\cite{2Arcizet2006, Corbitt2007, 2Corbitt2007, Caniard2007, MowLowry2008, Liu2021}, spanning from milligrams to grams. Consequently, gravitational measurements conducted in optomechanical systems in this mass scale range have become particularly of interest.

Currently, the proposed or conducted gravitational detection experiments have mainly been based on torsion balance setups~\cite{Balushi2018,Komori2018,Westphal2021}. In contrast, gravitational sensing that utilize microwave cavity optomechanical interactions and other types of mechanical resonators, such as pre-stressed membranes, remains less explored. These systems offer advantages such as higher mechanical quality factors~\cite{Fink2016, Schmid2016, Yuan2015, Lu2015, Torres2013, Sementilli2021, Liu2022, Pokharel2022, Seis2022} and reduced thermal noise~\cite{Fink2016, Teufel2008, Zhao2012, Pontin2016, Kalaee2019, Tang2023, Engelsen2024, Serra2021, liu2021optomechanical, Xu2024, Pokharel2022, Tsaturyan2017, Seis2022, Youssefi2023, Huang2024}, making them particularly favorable for gravitational detection. In this work, we present an optimized detection scheme using microwave circuit optomechanics to probe gravitational interactions between two milligram-scale test and source masses, building upon the established setup in~\cite{Liu2021}. This modulation is facilitated by a precisely controlled harmonic time-varying gravitational field, enhancing both interaction and measurement capabilities. Leveraging the properties of prestressed membranes, such as silicon carbide, which is known for its exceptional hardness which is second only to diamond, and a stress tolerance limit of up to 21~\si{GPa}~\cite{Erick2020,Sementilli2021}, allows the oscillators to support large enough gravitating masses while maintaining sufficiently high mechanical vibration frequencies~\cite{Liu2025}.

In contrast to~\cite{Liu2021}, where the gravitational interaction is probed via spectral measurements of the test mass’s motion, our scheme employs the linear response of the optomechanical induced transparency (OMIT)~\cite{Weis2010, Safavi2011, Xiong2012, Zhou2013, Xiong2018} to detect the resonant gravitational coupling. By driving the source mass, it undergoes periodic motion, generating a time-varying gravitational field that causes resonant motion in the oscillator loaded with the test mass. This motion of the test mass-loaded oscillator, in turn, leads to variations in the spectra of the OMIT, allowing us to identify the gravitational interaction between the test mass and the source mass. This method takes advantage of the sensitivity of the OMIT spectrum to weak dynamical forces in the setup, enabling us to discern the effects of gravitational interactions on the microwave optomechanical system. The proposed experimental scheme demonstrates the ability to measure exceptionally weak gravitational forces, reaching magnitudes as low as 6.4 attonewtons, thereby opening new avenues for detecting gravity induced by small mass objects within cavity optomechanical systems. Such efforts are part of a broader trend to extend laboratory tests of gravity toward ever smaller mass scales. Demonstrating gravity between milligram-scale resonators is an important step for isolating gravity as the dominant interaction in this regime, and provides a stepping stone toward future quantum experiments, ranging from entanglement-based probes of the quantization of gravity to the exploration of macroscopic quantum effects in larger mass mechanical systems under cavity optomechanics. In addition, some theoretical scenarios beyond the Standard Model predict deviations from Newtonian gravity at sub-millimeter scales. For example, Randall–Sundrum models with extra dimensions modify the Newtonian potential at distances up to about $0.169\ \si{mm}$~\cite{Buisseret2007, Pourhassan2022}. Since the spheres considered in our setup have radii of order $0.25\ \si{mm}$, our proposed scheme may also serve as a potential platform to test such models in future experiments. Furthermore, the use of milligram-scale oscillators provides additional opportunities to explore macroscopic quantum effects in larger mass mechanical systems under cavity optomechanics.

The structure of this paper is as follows: Section~\ref{sec:model} introduces the necessary models for discussion and presents the expression for the OMIT response in the presence of gravitational driving. Section~\ref{sec:results} assumes a set of plausible experimental parameters to illustrate the main phenomena and analyzes the effects of varying system parameters. Section~\ref{sec:DS} discusses the advantages of dynamic gravity sensing compared to static gravity sensing. Section~\ref{sec:sensitivity} discusses the sensitivity of the proposed system. Section~\ref{sec:conclusion} is the conclusion.

\section{Model}
\label{sec:model}
\subsection{Effective Hamiltonian for gravity interaction between two oscillators}
\begin{figure}[h]
    \centering    
    \hspace*{0cm} 
    \vspace*{-0.2cm}
    \includegraphics[width=1\linewidth]{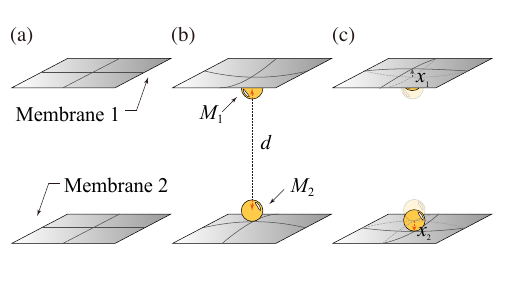}
    \caption{This figure illustrates the sphere-sphere gravitational interaction component in the setup under investigation. Fig.~\ref{fig:fig2}(a) shows the two membranes without the loaded masses. In Fig.~\ref{fig:fig2}(b), after loading the test mass $M_1$ and source mass $M_2$, the two membranes bend slightly towards each other due to gravitational attraction, with the equilibrium center of mass distance between the two masses being $d$. Fig.~\ref{fig:fig2}(c) depicts the small dynamic displacements $x_1$ and $x_2$ of the respective masses $M_1$ and $M_2$ away from equilibrium.
} 
    \label{fig:fig2}
\end{figure}

For two spheres with masses $M_1$ and $M_2$ separated by a center-of-mass distance $R$, it is reasonable to assume that gravity acts through Newtonian potential $V_G= -G M_1 M_2 / R$ where $G$ are the Newton constant. In the experimental setup under investigation, the core part are two mass loaded membranes, as shown in Fig.~\ref{fig:fig2}(a). After loading the test mass $M_1$ and source mass $M_2$ onto either membrane, the two membranes bend slightly towards each other due to their static gravitational attraction. At equilibrium, established by the combined influence of the ambient gravity and the spheres' gravity attraction, the center-of-mass distance between the two masses is $d$, as shown in Fig.~\ref{fig:fig2}(b). The membrane 2 loaded with the source mass $M_2$ is driven by a piezoelectric disk, generating a time-varying gravitational field that induces motion in membrane 1 loaded with test mass $M_1$. Consequently, one can express $R = d + x_2 - x_1$, where $d$ represents the distance between the masses at equilibrium, while $x_1$ and $x_2$ denote their respective displacements relative to the equilibrium position, as seen in Fig.~\ref{fig:fig2}(c). Since $\lvert x_2 - x_1 \rvert \ll d$, one could expand $V_G$ in terms of $(x_2 - x_1)/d$.

Following~\cite{Liu2021}, due to the suppression of higher-order terms by $(x_2 - x_1)/d$, it is reasonable to retain only quadratic terms involving $x_1$ and $x_2$. Since constant terms in the Hamiltonian do not affect the dynamics, and the linear terms in $x_1$ and $x_2$ vanish in equilibrium, the effective Hamiltonian for the gravitational interaction reduces to $H_G = -G M_1 M_2 (x_2 - x_1)^2 / d^3$.

\subsection{Optomechanical system driven by harmonic gravitational force}
\begin{figure}[htp]
    \centering    
    \hspace*{0cm} 
    \vspace*{0.4cm}
    \includegraphics[width=0.9\linewidth]{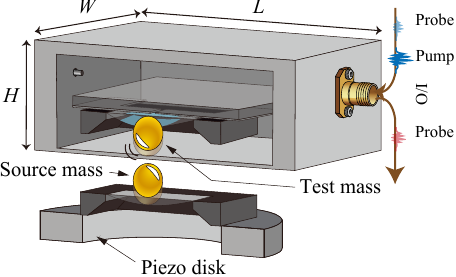}
    \caption{This figure shows the studied setup, which consists of an optomechanical cavity with length $L=48\ \si{mm}$, width $W=20\ \si{mm}$, and height $H=20\ \si{mm}$ with two coaxially aligned prestressed membranes (membrane~1 and membrane~2), loaded with a test mass $M_1$ and a source mass $M_2$, forming oscillators~1 and~2. These components are aligned along a parallel axis from top to bottom, with metalized membrane~1 and the cavity forming a 3D rectangular microwave cavity. Oscillator~2, consisting of mass $M_2$ loaded on membrane~2, is positioned on a piezoelectric disk. When an alternating voltage is applied, the disk drives oscillator~2 into vibration, causing $M_2$ to undergo harmonic motion with amplitude $x_s$ and frequency $\omega_s$. The input and output microwave ports are also indicated. Both the pump and probe fields are coupled through the same in/out SMA connector. A gray partition between the two spheres represents a conducting Faraday shield with thickness of $50\ \si{\mu m}$, which suppresses Coulomb interactions due to stray charges while leaving the modulated gravitational interaction unaffected.
} 
    \label{fig:fig3}
\end{figure}

As displayed in Fig.~\ref{fig:fig3}, the setup under investigation consists of two prestressed membranes 1 and 2, loaded with a test mass $M_1$ and a source mass $M_2$, forming mechanical oscillators denoted as oscillator~1 and~2. The natural frequencies of oscillators~1 and~2 are denoted as $\omega_1$ and $\omega_2$, with corresponding dissipation rates $\gamma_1$ and $\gamma_2$. The mechanical quality factors are $Q_1 = \omega_1 / \gamma_1$ and $Q_2 = \omega_2 / \gamma_2$, respectively. These components are aligned in parallel and coaxially. The cavity and the metalized membrane~1 together form a standard 3D rectangular microwave cavity optomechanical structure, with length $L=48\ \si{mm}$, width $W=20\ \si{mm}$, and height $H=20\ \si{mm}$, which is electromechanically coupled through an antenna electrode~\cite{Liu2022}. Oscillator~2, consisting of mass $M_2$ loaded on membrane~2, is positioned on a piezoelectric disk. When an alternating voltage is applied, the disk drives oscillator~2 into vibration, causing $M_2$ to undergo harmonic motion with amplitude $x_s$ and frequency $\omega_s$, described by $x_2 = x_s \cos (\omega_s t + \phi_s)$. Both the pump and probe microwave fields are coupled to the cavity through the same input/output (in/out) SMA connector, as indicated in Fig.~\ref{fig:fig3}. A gray partition with thickness of $50\ \si{\mu m}$ is inserted between the two masses, representing a conducting Faraday shield that suppresses stray-charge-induced Coulomb interactions while leaving the gravitational interaction unaffected. This motion causes the source mass $M_2$ to generate a periodically time-varying gravitational field. When the natural frequencies of the two oscillators are matched, the oscillator~1 exhibits the strongest response under this driving field. Further, this resonance leads to a variation in the microwave field within the microwave cavity, which is discernible by comparing the spectra in the OMIT configuration in the presence and absence of gravitational driving, the latter achieved by turning off the alternating voltage to make $x_s = 0$.

By replacing the position and momentum $x_1$ and $p_1$ of test mass oscillator to its corresponding quantum canonical operators $\hat{x}_1$ and $\hat{p}_1$, the above system is described as a cavity optomechanical system with time-harmonic driving force, characterized by the following Hamiltonian
\begin{equation}
\begin{aligned}
        \hat{H}_{tot}&=\frac{1}{2 M_1}\hat{p}_1^2 +\frac{1}{2}M_1 \omega_1^2 \hat{x}_1^2 + \hbar \omega_{c}\hat{a}^{\dagger}\hat{a} \\
        &- \hbar G_{\rm om}\hat{x}_1 \hat{a}^{\dagger}\hat{a} -\frac{G M_1 M_2}{d^3}[x_s \cos (\omega_s t + \phi_s) - \hat{x}_1]^2 \\
        &+ [i\hbar \sqrt{\eta_c \kappa}(\alpha_{l}e^{-i\omega_{l}t-i\phi_l}+\alpha_{p}e^{-i\omega_{p}t-i\phi_p})\hat{a}^{\dagger}+\text{h.c.}].
\end{aligned}
\end{equation}
Here, the first and second terms represent the test mass oscillator with frequency $\omega_1$ and effective mass $M_1$. The third term represents the cavity field with frequency $\omega_c$. The fourth term represents the optomechanical coupling, where $G_{om}$ is the optomechanical frequency shift per displacement. The fifth term represents the gravity coupling contributed by a source mass $M_2$, which undergoes harmonic motion with amplitude $x_s$, frequency $\omega_s$, and initial phase $\phi_s$, at a distance $d$ from $M_1$. The sixth term represents the pump and probe tone, where the amplitudes of the pump and probe tone are $\alpha_{l}$ and $\alpha_{p}$, with frequencies $\omega_{l}$ and $\omega_{p}$, and initial phases $\phi_{l}$ and $\phi_{p}$, respectively, while $\kappa = \kappa_{ext} + \kappa_{in}$ is the cavity dissipation and $\eta_c = \kappa_{ext}/\kappa$ is the coupling parameter with $\kappa_{ext}$ and $\kappa_{in}$ the external and intrinsic dissipation, respectively. 
    
The quantum expectation value $\langle \hat{x}_1 \rangle$ and $\langle \hat{a} \rangle$ of the system can be perturbatively found by taking the ansatz
\begin{equation}
    \begin{aligned}
        \langle \hat{x}_1 \rangle &= \overline{x}_1 + \delta x_1,\quad \langle \hat{a}\rangle &= \overline{a} + \delta a,
    \end{aligned}
\end{equation}
where $\overline{x}_1$ and $\overline{a}$ are the stationary solutions for $\langle \hat{x}_1 \rangle$ and $\langle \hat{a} \rangle$ while
\begin{equation}
    \begin{aligned}
        \delta x_1 =& A_1 P^- + A'_1 G^- + A^*_1 P^+ + A^*_1 G^+,\\
        \delta a =& e^{i\arg \overline{a}}(B P^- + B' G^- + C P^+ + C' G^+),
    \end{aligned}
\end{equation}
are the first-order respones due to the probe tone and gravitational driving with (The expressions for $A$, $B$, $C$, and more details, are given in Appendix~\ref{sec:cABC})
\begin{equation}
    \begin{aligned}
        P^\pm &= \sqrt{\eta_c \kappa}\alpha_{p}e^{\pm (i\omega t +i\phi_p +i\arg \overline{a})},\\
        G^\pm &= -\frac{2 G M_2 x_s}{d^3}e^{\pm (i\omega_s t + i\phi_s)}.
    \end{aligned}
\end{equation}
Here, $\Delta = \omega_l - \omega_c$ and $\omega = \omega_p - \omega_l$ represent the detuning between the pump field and cavity field as well as the detuning between the probe field and pump field. We also define the quantities: $\overline{\Delta} = \Delta + G_{om} \overline{x}_1$, which represents the effective detuning shifted by the mechanical displacement, and $\omega'_1 = \sqrt{\omega_1^2 - 2 G M_2 / d^3}$, which is the frequency shift caused by the gravitational influence of the source mass. In practical calculations, it can be assumed that $\overline{\Delta}\approx \Delta$ and $\omega'_1\approx \omega_1$.
    
\begin{figure}[h]
    \centering    
    \hspace*{0cm} 
    \vspace*{0cm}
    \includegraphics[width=\linewidth]{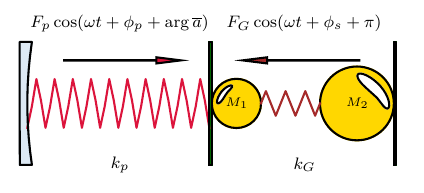}
    \caption{The effect of the cavity field's radiation pressure and the gravitational interaction of mass $M_2$ on oscillator 1. The radiation pressure from the cavity field acting on oscillator 1 is equivalent to a spring with spring constant $k_p$ (which equals zero when $\omega = -\overline{\Delta}$) and a periodic driving force $F_p \cos(\omega t + \phi_p + \arg\overline{a})$, where the former corresponds to the optical spring effect and the latter to the driving force induced by the probe tone $\alpha_p$. Similarly, the gravitational force from $M_2$ acting on oscillator 1 can be equivalently represented as a spring with spring constant $k_G$ and a periodic driving force $F_G \cos(\omega t + \phi_s + \pi)$, where the former corresponds to the spring effect due to the gravitational gradient of $M_2$ and the latter originates from the vibration of $M_2$.
} 
    \label{fig:fig4}
\end{figure}

In the following, OMIT of the system will be the main topic of discussion, therefore the quantity $\delta a$ will be of interest. Since the contributions from $C$ and $C'$, corresponding to the far-off-resonance lower sideband, are negligible, $\delta a$ depends primarily on $B$ and $B'$ at the considered frequency. Thus, we have:
\begin{equation}
    \begin{aligned}
        \delta a = e^{i\arg \overline{a}}(B P^- + B' G^-).
    \end{aligned}
\end{equation}
Furthermore, to achieve a stationary spectrum, it is necessary to ensure that $\omega_s = \omega$. With this condition  $\delta a$ becomes
\begin{equation}
    \begin{aligned}
        \delta a = \frac{1+i f(\omega)[1+ r \frac{2}{\kappa} e^{i\phi}(i\overline{\Delta}-i\omega + \frac{\kappa}{2})]}{-i(\overline{\Delta}+\omega)+\frac{\kappa}{2}+2\overline{\Delta}f(\omega)} \sqrt{\eta_c \kappa}\alpha_{p}e^{-i\omega t - i\phi_p},
    \end{aligned}
        \label{eq:Deltaa}
\end{equation}
where
\begin{equation}
    \begin{aligned}
 \chi (\omega) =& \frac{1}{M_1 ({\omega'_1}^2 - \omega^2 - i\omega \gamma_1)},\\
 f(\omega) =& \hbar G_{om}^2 |\overline{a}|^2 \frac{\chi (\omega)}{i(\overline{\Delta} - \omega) + \frac{\kappa}{2}},
    \end{aligned}
\end{equation}
and
\begin{equation}
    \begin{aligned}
        r =& \frac{F_G}{F_p},\ F_G = \frac{2 G M_1 M_2 x_s}{d^3},\  F_p = \frac{4\hbar G_{om}|\overline{a}|\sqrt{\eta_c \kappa}\alpha_p}{\kappa},\\
        \phi =& \phi_{fp} -\phi_{G},\ \phi_{fp} = \arg \overline{a} + \phi_p,\ \phi_{G} = \phi_s + \pi. \\
    \end{aligned}
    \label{eq:kphi}
\end{equation}
Here, $F_p$ and $F_G$ denote the driving force from the probe tone at frequency $\omega = -\overline{\Delta}$ and the amplitudes of the gravitational driving force from oscillator 2 while $\phi_{fp}$ and $\phi_{G}$ denote their initial phases. The coefficients $r$ and $\phi$ represent the ratio of amplitudes and the phase difference of the two forces as seen in Fig.~\ref{fig:fig4}
(For more details, refer to Appendix~\ref{sec:FpFG}).

The transmission $t_p$ of the probe tone is define by the ratio of the output field amplitudes $\alpha_{p}e^{-i\omega t - i\phi_p} - \sqrt{\eta_c \kappa}\delta \alpha$ to the input field amplitudes $\alpha_{p}e^{-i\omega t - i\phi_p}$ at the probe frequency $\omega$. In the presence of gravitational driving, it is given by (For more details, refer to Appendix~\ref{sec:ctpG})
\begin{equation}
    \begin{aligned}
        t_{pG} = 1 - \frac{1+i f(\omega)[1+re^{i\phi}\frac{2}{\kappa}(i\overline{\Delta}-i\omega + \frac{\kappa}{2})]}{-i(\overline{\Delta}+\omega)+\frac{\kappa}{2}+2\overline{\Delta}f(\omega)}\eta_c \kappa.
    \end{aligned}
    \label{eq:tpG}
\end{equation}
It can be observed that the parameter $r e^{i\phi}$ in Eq.~\eqref{eq:Deltaa} and Eq.~\eqref{eq:tpG} characterizes the impact of gravitational driving on the system, with $t_p$ reduced to its form
\begin{equation}
    \begin{aligned}
        t_{p0} = 1 - \frac{1+i f(\omega)}{-i(\overline{\Delta}+\omega)+\frac{\kappa}{2}+2\overline{\Delta}f(\omega)}\eta_c \kappa,
    \end{aligned}
    \label{eq:tp0}
\end{equation}
in the absence of gravitational driving when $r = 0$ ($x_s =0$). 

\section{Results}
\label{sec:results}
In this section, a set of  plausible experimental parameters for the system is presented to emphasize the main impact of gravitational driving on the spectra. Following this,
an examination will be conducted to assess how the transmssion spectra are impacted by different values of experimental parameters for the system.
\subsection{Plausible experimental parameters}
Central to the proposed setup are two membranes, each carrying a test mass and a source mass, loaded on two square membranes, e.g., made from silicon carbide (SiC). These membranes have side lengths $l=5 \ \si{mm}$, thickness $b=100\ \si{nm}$, and density $\rho_\text{SiC}=3.21\ \si{g\per cm^{3}}$. The test and source masses are gold spheres with density $\rho_\text{Au} = 19.3\ \si{g\per cm^{3}}$ and radii $r_1 = r_2 = 0.25 \ \si{mm}$, resulting in masses of $M_1 = M_2 = 1.26 \ \si{mg}$, with their centers of mass separated by a distance of $d=0.55\ \si{mm}$, which allows enough physical separation for e.g., a Faraday shield.
    
Before the masses are loaded, a transverse prestress of $\sigma = 10\ \si{GPa}$ is applied to each of the silicon carbide membranes. This results in vibration frequencies of $\omega_0 /2 \pi = \sqrt{\sigma /2\rho_\text{SiC} l^2} = 249.6\ \si{kHz}$. The high prestress effectively mitigates internal material losses, allowing for a possible quality factor of up to $Q_0 \lesssim 10^9$ especially at low temperature, e.g., $T = 10\ \si{mK}$.
    
After the masses are loaded, the vibration frequency and the quality faxctor of the membrane substantially decrease since the total mass and the total mechanical losses of the ocsillator increase. During the loading process, epoxy resin is used as glue to attach the gold spheres to the membrane. Once solidified, the epoxy resin is modeled as a  short cylinder with a radius of $r_\text{glue} = 100\ \si{\mu m}$ and a height of $h = 20\ \si{\mu m}$.
    
The finite element method is employed to simulate the silicon carbide membranes loaded with gold spheres as described above, yielding their eigenfrequencies. The simulation results reveal that when the oscillator possesses the aforementioned parameters, its lowest eigenfrequency is approximately $\omega_1 /2\pi = 8\ \si{kHz}$ while the quality factor decreases to $Q_1 = \omega_1 / \gamma_1 \sim 10^7$. The cavity field dissipation is set to be $\kappa = \omega_1 = 2\pi\times 8\ \si{kHz}$ with the critical coupling $\eta_c = 1/2$. Meanwhile, the pump tone is tuned to the lower motional sideband $\overline{\Delta} = -\omega'_1 \approx -2\pi\times 8\ \si{kHz}$.
    
The optomechanical frequency shift per displacement is taken to be $G_{om} = 5\times 10^{15}\ \si{Hz/m}$ similar to~\cite{Liu2021} while the average photon number is $|\overline{a}|^2 = 10^4$ which corrosponding to an enhanced, laser-tunable optomechanical coupling strength $g = g_0 |\overline{a}| = 2\pi\times 2.3\ \si{Hz}$. Here $g_0 = G_{om} x_\text{ZPF1}$ is the coupling between a single photon and a single phonon while $x_\text{ZPF1} = \sqrt{\hbar / 2 M_1 \omega'_1} = 29\ \si{am}$ is the zero point fluctuation of oscillator 1.
    
The power of the probe tone is chosen to be $P_{\text{prob}} \sim 1\ \si{aW}$, operating at frequency e.g. $\omega_\text{prob} / 2\pi = 5\ \si{GHz}$. Under this probe tone power, the average probe photon number is given by
\begin{equation}
    \begin{aligned}
        \overline{n}_p = \eta_c \frac{P_p}{\hbar \omega_p} \frac{4\kappa}{\kappa^2 + 4(\omega_p - \omega_c)^2} \sim 10^2,
    \end{aligned}
\end{equation}
corresponding to a probe tone signal-to-noise ratio ${\rm SNR}_{\rm shot} = \sqrt{\overline{n}_p} \sim 10$ when $\omega_p = \omega_c$. By appropriately adjusting the driving alternating voltage, it is plausible to reach the driven motion $x_s = 5\ \si{\mu m}$ of the source mass $M_2$ when it is driven with a frequency equal to $\omega'_1$. Since $\phi = \arg \overline{a}+\phi_p -\phi_s -\pi, \overline{a} = \sqrt{\eta_c \kappa}\alpha_{l}e^{-i\phi_l}/(\kappa /2-i\overline{\Delta})$ while $\phi_{l}$, $\phi_{p}$, $\phi_{s}$ are adjustable, it is possible to set $\phi$ to any value between $-\pi$ and $\pi$. Here, $\arg \overline{a}, \phi_p, \phi_s$ is chosen to be $0,0,-\pi$ respectively. Under these values of parameters, the amplitude of gravitational driving force and probe tone driving force are $F_G = 6.4\ \si{aN}$ and $F_p = 365.5\ \si{aN}$, with the ratio $r = F_G / F_p = 1.75\times 10^{-2}$.
    
\begin{figure}[h]
    \centering    
    \hspace*{-0.7cm} 
    \vspace*{-0.5cm}
    \includegraphics[width=\linewidth]{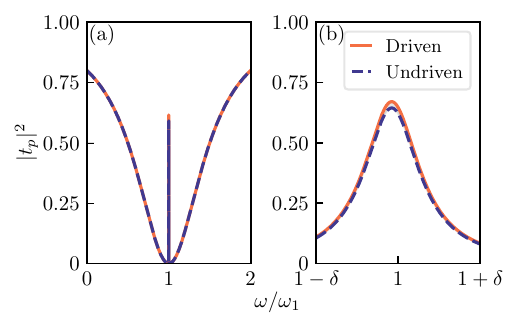}
    \caption{The figures demonstrate the transmission spectra $|t_p|^2$ in the presence (red line, labelled as 'Driven') and absence of gravitational driving (blue dashed line, labelled as 'Unriven') under plausible experimental parameters from Table~\ref{tab:table1}. Fig.~\ref{fig:fig5}(a) demonstrates the overall transmission spectra for both cases. In Fig.~\ref{fig:fig5}(b), the spectra near $\omega = \omega'_1$ are magnified by a factor $\delta = 5\times 10^{-7}$ in the $x$-direction.}
    \label{fig:fig5}
\end{figure}

Based on the above data, plausible experimental parameters describing the system can be obtained, as provided in Table~\ref{tab:table1}. 
\begin{table}[b]
\caption{The plausible experimental parameters used in estimating the transmission coefficients}
\label{tab:table1}
\begin{ruledtabular}
\begin{tabular}{ccc}
Symbol & Description & Value\\
\hline
$T$ & Equilibrium mode temperature & $10\ \si{mK}$ \\
$M_1$ & Test mass & $1.26\ \si{mg}$ \\
$M_2$ & Source mass & $1.26\ \si{mg}$ \\
$d$ & C.o.M. distance between $M_1$ and $M_2$ & $0.55\ \si{mm}$ \\
$\omega_1$ & Natural frequency of oscillator 1 & $2\pi\times 8\ \si{kHz}$\\
$\gamma_1$ & Dissipation of oscillator 1 & $2\pi\times 0.8\ \si{mHz}$ \\
$Q_1$ & $Q$ value of oscillator 1 & $10^7$\\
$x_s$ & Amplitude of oscillator 2 & $5\ \si{\mu m}$ \\
$\phi_s$ & Initial phase of oscillator 2& $-\pi$ \\
$P_p$ & Power of probe tone & $10^{-18} \ \si{W}$ \\
$\omega_p$ & Frequency of probe tone & $2\pi\times 5\ \si{GHz}$ \\
$\phi_p$ & Initial phase of probe tone & $0$ \\
$|\overline{a}|$ & Amplitude of average cavity field at $\omega =\omega_1$ & $10^2$ \\
$\arg \overline{a}$ & Phase of average cavity field & $0$ \\
$\Delta$ & Detune $\omega_l -\omega_c$ & $2\pi\times 8\ \si{kHz}$ \\
$\kappa$ & Cavity dissipation & $2\pi\times 8\ \si{kHz}$ \\
$\eta_c$ & Coupling parameter $\kappa_{ext} / \kappa$ & $0.5$ \\
$g$ & Enhanced optomechanical coupling & $2\pi \times 2.3\ \si{Hz}$ \\
$F_G$ & Amplitude of gravitational driving force & $6.4\ \si{aN}$ \\
$F_p$ & Amplitude of probe tone driving force & $365.5\ \si{aN}$ \\
$r$ & $F_G / F_p$ & $1.75\times 10^{-2}$\\
$\phi$ & $\arg \overline{a} + \phi_p -\phi_s -\pi$ & $0$ \\
\end{tabular}
\end{ruledtabular}
\end{table}
Under these parameters, transmission spectra $|t_p|^2$ of the system can be plotted, as shown in Fig.~\ref{fig:fig5}, where $\Delta |t_p|^2 = |t_{pG}|^2 - |t_{p0}|^2$ is introduced to characterize the change in the transmission due to gravitational driving. As can be seen in Fig.~\ref{fig:fig5}, gravitational driving induces variation $\Delta |t_p|^2 = |t_{p0}|^2 - |t_{pG}|^2$ in the transmission rate $|t_p|^2$. Notably, the transmission spectrum experiences significant variations, particularly near $\omega = \omega'_1$, where, under plausible experimental parameters, the maximum variation in $|t_{pG}|^2$ in the presence of gravity compared to its absence $|t_{p0}|^2$ can reach up to $\Delta |t_p|^2_{\rm max} \approx 0.023$. Here, the deviation due to the shot noise of the probe tone is considered.

Following~\cite{Liu2021}, the measurement time $\tau$ for a single datapoint along the spectral curves in Fig.~\ref{fig:fig5} is given by:

\begin{equation}
    \tau = \frac{F_G^2}{S^{\rm eff}_f (\omega)},
\end{equation}
where $S^{\rm eff}_f$ is the effective force noise spectrum, defined as

\begin{equation}
    S^{\rm eff}_f = S^{\rm ZP}_f + S^T_f + S^E_f + S^{\rm qba}_f + S^{\rm imp}_f.
\end{equation}
The terms $S^{\rm ZP}_f$, $S^T_f$, $S^E_f$, $S^{\rm qba}_f$, and $S^{\rm imp}_f$ represent the contributions from zero-point mechanical noise, thermal noise, external vibration noise, quantum backaction noise, and imprecision noise, respectively, given by

\begin{equation}
    \begin{aligned}
    S^{\rm ZP}_f (\omega) &= \gamma_1 \hbar M_1 \omega_1, \\
    S^T_f (\omega) &= 2 \gamma_1 \hbar M_1 \omega_1 \overline{n}_1, \\
    S^E_f (\omega) &= M_1^2 \omega_1^4 S^E_x (\omega), \\
    S^{\rm qba} (\omega) &= \frac{2\mathcal{C}}{x_\text{ZPF1}^2} \gamma \hbar^2, \\
    S^{\rm imp}_f (\omega) &= \frac{1/2 + S_{\rm add}}{4\mathcal{C} \gamma_1 |\chi (\omega)|^2 / x_\text{ZPF1}^2}.
    \end{aligned}
\end{equation}
Here, $\overline{n}_1 = 1 / [\exp(\hbar \omega_1 / k_B T) - 1]$ and $\mathcal{C} = 4 g^2 / (\kappa \gamma_1)$ are the equilibrium phonon number and the cooperativity of oscillator 1, respectively. Using the experimental parameters in Table~\ref{tab:table1}, $\tau$ is estimated to be approximately $1.9 \times 10^3 \ \text{s}$.

It is important to note that the ratio of the frequency resolution required to resolve the OMIT peak to the mechanical frequency is approximately given by $1/Q_1$. For the parameters listed in Table~\ref{tab:table1}, this corresponds to a ratio of $10^{-7}$, which translates to a frequency resolution on the order of $\si{mHz}$. Such high resolution can in practice be achieved by extending the sampling time and performing FFT on the time-domain signal. Indeed, frequency resolutions exceeding this level have already been demonstrated in a similar optomechanical system~\cite{Liu2025}, featuring a mechanical frequency of $750\ \si{kHz}$ and a frequency resolution below $1\ \si{mHz}$ to assess the linewidth, correspond to a ratio of $1.3\times 10^{-9}$. 

\subsection{Impact of Experimental Parameters on the Transmission Spectrum}
In this subsection, the impact of different values of system parameters on the transmission spectrum will be discussed. It can be observed that the parameters $\omega'_1$, $Q_1$, $\eta_c$, $\kappa$, $\overline{\Delta}$, $g$, $r$, and $\phi$ fully determine the system. To simplify the discussion on the impact of these parameters on the spectra, $\eta_c$ is maintained at critical coupling $\eta_c = 1/2$ while the pump tone is tuned to the lower motional sideband $\overline{\Delta} = -\omega'_1$. 

Worth noting that the five parameters above can be divided into two groups. One group includes $\omega'_1$, $\kappa$, $g$, and $Q_1$, which are typically required to describe an optomechanical system. When these four parameters vary, the $|t_p|^2$ curves in the presence and absence of gravitational drive will change. The other group includes $r$ and $\phi$, which are parameters related to the gravitational driving. When these two parameters vary, only the $|t_p|^2$ curve in the presence of gravitational drive will vary.

To intuitively understand the impact of the variation of $|t_p|^2$ near the transmission windows with respect to different system parameters mentioned above, it is worth noting that as $\Delta' = \omega - \omega'_1 \rightarrow 0$, the expressions of $t_{pG}$ and $t_{p0}$ in Eq.~\eqref{eq:tpG} and \eqref{eq:tp0} can be approximated by standard Lorentzian functions
\begin{equation}
    \begin{aligned}
        t_{pG} \approx \frac{4g^2(1+i\frac{\kappa}{4\omega'_1})(1+re^{i\phi})}{4g^2 + \kappa \gamma_1 + \frac{\kappa^2}{2\omega'_1}\Delta' +i(\frac{\kappa^2 \gamma_1}{4\omega'_1} -2\kappa\Delta')},
    \end{aligned}
    \label{eq:tpG2approx2}
\end{equation}
and
\begin{equation}
    \begin{aligned}
        t_{p0} \approx \frac{4g^2(1+i\frac{\kappa}{4\omega'_1})}{4g^2 + \kappa \gamma_1 + \frac{\kappa^2}{2\omega'_1}\Delta' +i(\frac{\kappa^2 \gamma_1}{4\omega'_1} -2\kappa\Delta')},
    \end{aligned}
    \label{eq:tp02approx2}
\end{equation}
with the conditions $\eta_c =1/2$ and $\overline{\Delta} = -\omega'_1$ (For more details, refer to Appendix~\ref{sec:tpGrw}).
\begin{figure}[htp]
    \centering    
    \hspace*{-0.7cm} 
    \vspace*{0cm}
    \includegraphics[width=\linewidth]{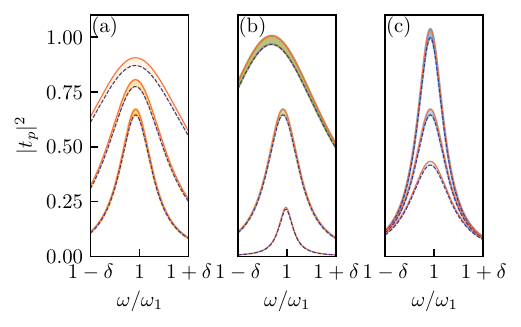}
    \caption{This figure illustrate the effects of varying $\kappa$, $g$, and $Q_1$ around their reference values on the transmission spectra, with $r$ and $\phi$ fixed. Solid and dashed lines correspond to the presence and absence of gravitational effects, respectively. In Fig.~\ref{fig:fig6}(a), the spectra for $\kappa$ at the reference value, half, and a quarter of the reference value are shown with progressively lighter fill colors. In Fig.~\ref{fig:fig6}(b), the spectra for $g$ at twice, the reference value, and half are shown similarly. In Fig.~\ref{fig:fig6}(c), the spectra for $Q_1$ at five times, the reference value, and half are shown in the same manner.}
    \label{fig:fig6}
\end{figure}

As $\Delta' \rightarrow 0$, Eq.~\eqref{eq:tpG2approx2} and \eqref{eq:tp02approx2} indicate that the transmission $|t_{pG}|^2$ is $|t_{p0}|^2 |1 + re^{i\phi}|^2$, and the phase dispersion $\arg t_{pG}$ is $\arg t_{p0} + \arg(1 + re^{i\phi})$. Here, $|t_{p0}|^2$ and $\arg t_{p0}$ depend solely on the optomechanical parameters $\kappa$, $g$, and $Q_1$, while $|1 + re^{i\phi}|^2$ and $\arg(1 + re^{i\phi})$ are determined by the gravitational driving parameters $r$ and $\phi$. Thus, for two systems with identical optomechanical parameters $\omega'_1$, $\kappa$, $Q_1$, and $g$, one in the presence of gravitational driving and the other in its absence, the peak height differs only by a factor of $|1 + re^{i\phi}|^2$, and the phase dispersion is globally shifted by $\arg(1 + re^{i\phi})$. Meanwhile, the peak width and position remain identical.

As shown in the Fig.~\ref{fig:fig6}, varying the optomechanical parameters $\kappa$, $g$, and $Q_1$ individually around a set of reference values of $\omega'_1 \approx \omega_1 = \kappa = 2\pi \times 8 \ \si{kHz}$, $g =2\pi \times 2.3~\si{Hz}$, $Q_1 = 10^7$, $r = 1.75\times 10^{-2}$ and $\phi =0$ according to Table ~\ref{tab:table1} results in simultaneous changes in the transmission spectra in the presence and absence of gravitational driving, with the peak height changing proportionally. The peak width and position also shift, but these changes occur synchronously; for the same set of optomechanical parameters, the peak width and position in the presence and absence of gravitational driving remain identical, in agreement with the previous analysis.

\begin{figure}[h]
    \centering    
    \hspace*{-1.5cm} 
    \vspace*{0cm}
    \includegraphics[width=\linewidth]{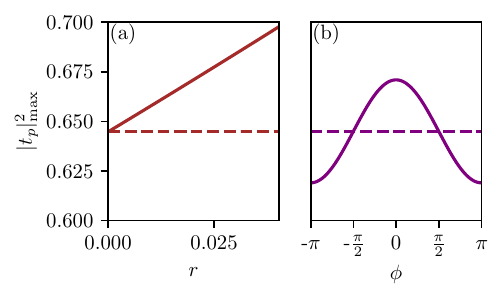}
    \caption{This figure illustrates the variation in the peak height of the transmission spectra as $r$ and $\phi$ are varied independently. The solid lines represent the case in the presence gravitational driving, while the dashed lines correspond to the case in the absence gravitational driving. Fig.~\ref{fig:fig7}(a) shows the dependence of the peak height on $r$, and Fig.~\ref{fig:fig7}(b) shows the dependence of the peak height on $\phi$.}
    \label{fig:fig7}
\end{figure}

When the optomechanical parameters are fixed at their reference values, varying $r$ and $\phi$ independently results in a change in $\Delta |t_p|^2_{\max}$ that is approximately proportional to $r$ and $\cos\phi$ as seen in Fig.~\ref{fig:fig7}(a) and (b), respectively. Specifically, for a fixed $r$, when $\phi = 0$, $\Delta |t_p|^2 \approx 2r$ reaches its positive maximum, while when $\phi = \pm \pi$, $\Delta |t_p|^2 \approx -2r$ reaches its negative maximum. This follows from $r \ll 1$, where $\Delta |t_p|^2_{\max} = (|1 + r e^{i\phi}|^2 - 1) |t_{p0}|^2 \approx 2r \cos\phi |t_{p0}|^2$.

\begin{figure}[h]
    \centering    
    \hspace*{-1cm} 
    \vspace*{0cm}
    \includegraphics[width=0.9\linewidth]{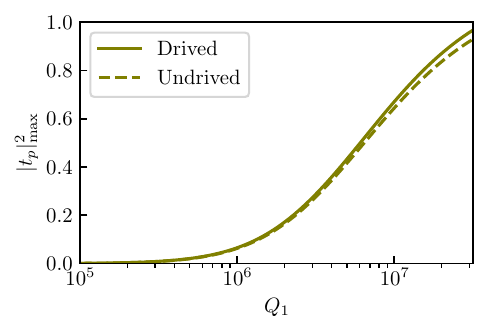}
    \caption{Transmission spectra as a function of $Q_1$. Solid and dashed lines represent spectra in the presence and absence of gravitational driving, respectively.}
    \label{fig:fig8}
\end{figure}

Furthermore, it is particularly noteworthy that the value of $\Delta |t_p|^2_\mathrm{max}$ is highly sensitive to the order of magnitude of $Q_1$. Specifically, when $Q_1 \lesssim 10^6$ is relatively small, $\Delta |t_p|^2_\mathrm{max}$ becomes very small and difficult to distinguish, as shown in Fig.\ref{fig:fig8}.

\subsection{Effects of Casimir and Electrostatic Forces}

In addition to gravity, the OMIT resonance can also be influenced by the Casimir interaction and stray electrostatic charges. Following Ref.~\cite{Plato2023}, the relative importance of these effects is quantified by three parameters: $\gamma_G$, $\gamma_C$, and $\gamma_Q$, which represent the strengths of the gravitational, Casimir, and electrostatic forces, respectively. These parameters are defined as
$\gamma_G = G M / d^3 \omega_m^2$,
$\gamma_C = 161 \hbar c R^6 / \pi M \omega_m^2 d^9$, and
$\gamma_Q = Q_e^2 / 4 \pi \epsilon_0 M d^3 \omega_m^2$.
Here, $M$ denotes the source and test mass (assuming $M_1 = M_2 = M$), $R$ the sphere radius, $d$ the center-of-mass distance between the two masses, and $Q_e$ the net stray charge on each mass. To clearly distinguish the gravitational contribution, one requires $\gamma_C, \gamma_Q \ll \gamma_G$.

For the parameters considered in Table~\ref{tab:table1}, the condition $\gamma_C \ll \gamma_G$ amounts to $M \gg 20~\mu\mathrm{g}$, which is comfortably satisfied since the assumed source mass is in the milligram range. By contrast, the requirement $\gamma_Q \ll \gamma_G$ demands that the net stray charges on each mass be less than $\sim 200\,e$, which is experimentally challenging. However, this constraint can be significantly relaxed by introducing a conducting Faraday shield between the two masses, which suppresses electrostatic interactions without affecting the modulated gravitational signal.

\section{Comparison with dynamic and static Gravity Sensing}
\label{sec:DS}
In the previous sections, dynamic gravity sensing is employed by comparing the variations in the OMIT spectra caused by the gravitational effect of $M_2$ on $M_1$ when the second oscillator is driven and not driven (with $M_2$ loaded on the second membrane in both cases). Compared to static gravity sensing, which senses gravity by comparing the variations in the OMIT spectra when the second membrane is loaded (but not driven) and not loaded with $M_2$, the advantages are as follows.
    
For dynamic gravity sensing, the transmission $|t_p|^2$ in the presence and absence of gravitational driving is given by the squared norm of Eq.~\eqref{eq:tpG} and~\eqref{eq:tp0}, respectively. The differences of peak height, peak position, and peak width in the presence of gravitational driving compared to in the absence of gravitational driving can be obtained through direct numerical calculations under plausible experimental parameters, given by 
\begin{equation}
    \begin{aligned}
        \Delta |t_p|^2  &= 2.3\times 10^{-2},\\
        \Delta \omega_{\rm max} &= -1.9\times 10^{-11}\ \si{Hz},\\
        \Delta \Gamma &= 3.6\times 10^{-10}\ \si{Hz}.\\
    \end{aligned}
    \label{eq:drivenvsundriven}
\end{equation}
For the static gravity sensing, when $M_2$ is loaded on the second membrane but not driven, the situation is exactly the same as the undriven case in dynamic sensing, i.e., $t_{p,{\rm loaded}} = t_{p0}$. When $M_2$ is unloaded, the only change is the disappearance of the static gravitational effect on the natural frequency of oscillator 1, i.e., $\omega'_1 = \sqrt{\omega_1^2 - 2 G M_2 / d^3}$ reverts back to $\omega_1$. In this case, $t_{p,{\rm unloaded}}$ is given by
\begin{equation}
    \begin{aligned}
        t_{p0} = 1 - \frac{1+i f(\omega)}{-i(\overline{\Delta}+\omega)+\frac{\kappa}{2}+2\overline{\Delta}f(\omega)}\eta_c \kappa,
    \end{aligned}
\end{equation}
where
\begin{equation}
    \begin{aligned}
 \chi (\omega) =& \frac{1}{M_1 ({\omega_1}^2 - \omega^2 - i\omega \gamma_1)},\\
 f(\omega) =& \hbar G_{om}^2 |\overline{a}|^2 \frac{\chi (\omega)}{i(\overline{\Delta} - \omega) + \frac{\kappa}{2}}.
    \end{aligned}
\end{equation}
Under plausible experimental parameters, the differences in peak height, position, and peak width when $M_2$ is loaded compared to when $M_2$ is unloaded can be numerically determined, given by:
\begin{equation}
    \begin{aligned}
        \Delta |t_p|^2  &= 1.1\times 10^{-16},\\
        \Delta \omega_{\rm max} &= -1.0\times 10^{-11}\ \si{Hz},\\
        \Delta \Gamma &= 5.2\times 10^{-18}\ \si{Hz}.\\
    \end{aligned}
    \label{eq:loadvsunload}
\end{equation}
From Eq.~\eqref{eq:drivenvsundriven} and \eqref{eq:loadvsunload}, it is evident that compared to static gravity sensing, dynamic gravity sensing results in a much more pronounced variation in peak heights when gravitational driving is present compared to its absence. 

Physically, this drastic difference between dynamic and static sensing originates from the distinct mechanisms by which dynamic and static gravity modify the OMIT spectrum. In the dynamic case, the oscillating gravitational field acts as an additional coherent drive for the mechanical resonator. This effectively adds a second excitation path to the phonon channel. When this external drive is in phase with the mechanical motion ($\phi = 0$), the phonon response is enhanced, leading to a stronger indirect channel and more complete destructive interference with the direct optical absorption. Consequently, the OMIT peak height increases. When the drives are out of phase ($\phi = \pi$), the phonon response is suppressed, and the transparency window becomes shallower. Quantitatively, the transmission follows a relative variation $|1+r e^{i\phi}|^2 - 1$, with $r = F_G /F_p$ the ratio between gravitational and probe driving forces according to Eq.~\eqref{eq:tpG2approx2} and \eqref{eq:tp02approx2}. Since the probe can be extremely weak, the ratio $r$ could reach up to a magnitude of $10^{-2}$ in our setup, resulting a relative variation of the the transmission to the magnitude of $10^{-2}$.

In contrast, in the static case, the static gravitational field does not provide a coherent drive but only shifts the natural frequency of the resonator, namely $\omega_1 \to {\omega'}_1 = \sqrt{\omega_1^2 - 2GM/d^3}$. To see its impact on the transmission, one could let $\Delta' =0$ in the Eq.\eqref{eq:tp02approx2} to obtain the expression of the height of the transimission
\begin{equation}
    \begin{split}
        |t_{p0,\text{loaded}}|^2_\text{max} = |t_{p0}|^2_\text{max} = \frac{\mathcal{C}^2 (1 + \kappa^2 / 16{\omega'}_1^2)}{(\mathcal{C} + 1)^2 + \kappa^2 / 16{\omega'}_1^2}.
    \end{split}
    \label{eq:tp0height}
\end{equation}
Here, we do not strictly assume $\kappa \ll {\omega'}_1$, so that the Eq.~(\ref{eq:tp0height}) retains dependence on $\kappa / {\omega'}_1$. Replace ${\omega'}_1$ with $\omega_1$ representing the case where the source mass $M_2$ is unloaded, this gives 
\begin{equation}
    \begin{split}
        |t_{p0,\text{unloaded}}|^2_\text{max} = \frac{\mathcal{C}^2 (1 + \kappa^2 / 16\omega_1^2)}{(\mathcal{C} + 1)^2 + \kappa^2 / 16\omega_1^2}.
    \end{split}
\end{equation}
The relative variation of the transmission for the static case is then given by
\begin{equation}
    \begin{split}
        \frac{|t_{p0,\text{loaded}}|^2_\text{max} - |t_{p0,\text{unloaded}}|^2_\text{max}}{|t_{p0,\text{unloaded}}|^2_\text{max}} \approx K\frac{\Delta\omega_1}{\omega_1},
    \end{split}
\end{equation}
where 
\begin{equation}
    \begin{split}
        K = \frac{32\mathcal{C}(\mathcal{C} + 2)[(\kappa / 4\omega_1)^2]}{[1+(\kappa / 4\omega_1)^2][(\mathcal{C} + 1)^2 + (\kappa / 4\omega_1)^2]}
    \end{split}
\end{equation}
is a coefficient of order unity for the plausible parameters given in Table~\ref{tab:table1}, and $\Delta\omega_1 / \omega_1= ({\omega'}_1 - \omega_1)/\omega_1 \approx G M_2 / \omega_1^2 d^3$ denotes the relative frequency shift. Since this shift scales inversely with the square of the frequency, it is of the order of $10^{-16}$ for our parameters, and therefore the relative variation of the transmission is of the same order.

Specifically, this variation in the peak heights of the transmission spectra can be as large as $\Delta |t_p|^2_{\rm max} \approx 0.023$, which is observable within the precision limits of current instrumentation.

\section{Detection sensitivity of the system}

The sensitivity of a measurement scheme is generally defined as the response of the chosen readout observable $R$ to variations of the input physical quantity $Q$, namely $S = |\delta R/\delta Q|$. In the present context, the target quantity is the gravitational driving force between the source and test masses, $Q = F_G$, while the readout $R$ is given by the variation of numbers of the output probe photons $\Delta \overline{n}_p$ which is the input probe photons $\overline{n}_p$ for the undriven case times the variation of the OMIT transmission peak height between the driven and undriven cases,
\begin{equation}
    \begin{split}
        \Delta \overline{n}_p &=\overline{n}_p \Delta |t_p|^2_{\max}\\ 
        &=\overline{n}_p (|t_{pG}|^2_{\max} - |t_{p0}|^2_{\max})\\
        &=\overline{n}_p (|1 + re^{i\phi}|^2 - 1)\frac{\mathcal{C}^2 (1 + \kappa^2 /16{\omega'}_1^2}{(\mathcal{C} + 1)^2 + \kappa^2 /16{\omega'}_1^2}.
    \end{split}
\end{equation}

Since $r = F_G / F_p$, the sensitivity of our proposed scheme is
\begin{equation}
    \begin{split}
        S = |\frac{\delta \Delta \overline{n}_p}{\delta F_G}| = |\frac{2 (\cos\phi + r)\overline{n}_p}{F_p}\frac{\mathcal{C}^2 (1 + \kappa^2 /16{\omega'}_1^2)}{(\mathcal{C} + 1)^2 + \kappa^2 /16{\omega'}_1^2}|.
    \end{split}
\end{equation}

Substituting the parameters listed in Table~\ref{tab:table1}, we find that the present scheme achieves a sensitivity of $S \sim 10^{18}\ \si{N^{-1}}$. This can be further enhanced by increase pump powers, since $F_p \propto \sqrt{\overline{n}_p}$, it results that $S \propto \sqrt{\overline{n}_p}$. For example if probe power increase to $100\ \si{aW}$, the sensitivity increase to $S \sim 10^{19}\ \si{N^{-1}}$.

Following Ref.~\cite{Liu2021}, the sensitivity $S$ can be derived by combining the system's response, $\delta x (\omega) = \chi (\omega) \delta \mathrm{d} F_G$, with the output field relation, $\delta y_\text{out} (\omega) = - 2\sqrt{\mathcal{C}\gamma_1}/x_\text{ZPF1} \delta x (\omega)$. This leads directly to $S = |\delta y_\text{out} / \delta \mathrm{d} F_G| = |2\sqrt{\mathcal{C}\gamma_1}\chi (\omega)/x_\text{ZPF1}|$, which is independent of the number of probe photons, and for our parameters, it is approximately $\sim 10^{19}\ \si{N^{-1}}$.

For other schemes, it is more difficult to compare the sensitivity because different schemes use different readout observables that are not directly convertible. For example, in Ref.~\cite{Westphal2021}, gravity is measured by detecting variation in the peak height of the test mass oscillator's displacement power spectrum at the resonance frequency, which cannot be directly converted into photon numbers.

We now considering the mass scaling of the observable. It can be deduced by using the equation $F_G = 2GM^2 x_s / d^3$ with all other parameters fixed and setting $\phi=0$. This leads to the conclusion that $r \propto M^2$, and we have
\begin{equation}
    \begin{split}
        \Delta |t_p|^2_{\max} \propto (1+r^2)-1 = 2r+r^2.
    \end{split}
\end{equation}
Therefore, for small masses with $r\ll1$, corresponding to $M \ll 10\ \si{mg}$ for our system parameters, the signal scales quadratically as $\Delta |t_p|^2_{\max}\propto M^2$. While for larger masses ($r\gtrsim 1$), it exhibits a crossover to quartic scaling $\Delta |t_p|^2_{\max}\propto M^4$. This scaling behavior highlights the rapid growth of the observable with mass, and provides useful guidance for designing experiments that optimize the mass regime of interest.
\label{sec:sensitivity}

\section{Conclusion}
In this study, we introduced a mass-loaded microwave cavity optomechanical setup capable of modulating and transducing gravitational interactions between milligram-scale test masses and a driven source mass. By analyzing the transmitted signal $t_p$ in the presence of gravitational driving and comparing it with the transmission $t_p$ in the absence gravitational driving, it was observed that the influence of gravitational driving on $|t_p|^2$ is primarily reflected in the peak height, while the position $\omega_{\rm max}$ and width $\Gamma$ of the transparency window remain nearly unchanged. Specifically, the relative variation in the OMIT peak height is given by $|1+re^{i\phi}|^2-1$, where $r$ is the ratio of the amplitude of the gravitational driving force to the amplitude of the radiation pressure from the probe tone, and $\phi$ is their phase difference. Under plausible experimental parameters, this variation can reach up to 2.3\%, which is significant given current experimental conditions and holds substantial promise for detection. The sensitivity analysis shows that the system can achieve a force responsivity on the order of $10^{18} \sim 10^{19}\ \si{N^{-1}}$, with further improvement possible at higher probe power. This quantitative benchmark demonstrates that the OMIT-based approach is competitive with existing methods, and clarifies how the observable scales with the mass of the test bodies. 

These findings underscore the subtle role of gravitational interactions in optomechanical systems, thereby providing valuable insights to guide future experimental investigations into weak gravity detection. The milligram-scale mass of the system represents a new mass scale not yet explored in existing precision measurements of gravity, offering an opportunity to test Newton's law of gravity at small mass scales and even opens avenues for investigating the nuanced effects of gravitational interactions in quantum systems. In addition, the milligram-scale mass of the system represents a new mass scale not yet explored in existing precision measurements of gravity. This scale is particularly advantageous: it is large enough to suppress Casimir backgrounds that dominate at masses below $20\ \si{\mu g}$, thereby facilitating clearer isolation of gravitational coupling—an essential requirement for future experiments aiming to observe gravity-induced entanglement. At the same time, the sub-millimeter dimensions of the source and test masses approach the characteristic length scale predicted by certain beyond-Standard-Model theories such as the Randall–Sundrum model with extra dimensions, making the setup a promising platform for probing possible deviations from Newtonian gravity and testing the existence of extra dimensions.
\label{sec:conclusion}

\begin{acknowledgments}
The authors greatly thank the constructive discussions with Prof. M. A. Sillanp\"a\"a. We also thank L. Zheng for preparing Figure~\ref{fig:fig3}. The authors acknowledge the support of National Natural Science Foundation of China (No. 12374325, No. 92365210, No. 12304387, and No. 22303005). This work is also supported by Beijing Natural Science Foundation (Z240007), Beijing Municipal Science and Technology Commission (Grant No.~Z221100002722011). Y. Liu acknowledges the support of Young Elite Scientists Sponsorship Program by CAST (Grant No. 2023QNRC001).
\end{acknowledgments}

\clearpage
\appendix
\renewcommand{\thefigure}{\Alph{section}\arabic{figure}}
\renewcommand{\thetable}{\Alph{section}\arabic{table}}
\setcounter{table}{0} 

\onecolumngrid
\section{Tables}\label{sec:tablesGandOPTO}
\setcounter{figure}{0} 

\begin{table*}[b]
\vspace*{-1.5\baselineskip}
\caption{\label{tab:tableG} Some key experiments for gravity measurement between 1790 and 2024, involving various materials and geometries of mass sources, and different measurement methods. Each row includes the author(s) of the experiment, the value of the source mass and its material, the geometry, the measured value of the gravitational constant $G$, the relative uncertainty $\Delta G / G$ (in ppm), the year of the experiment, and the reference.}
\begin{ruledtabular}
\begin{tabular}{cllcccccc}
 & Author(s) & Source (total,kg) & Material & Geometry & $G ( 10^{-11}\ \si{m}^{3}\ \si{s}^{-2}\ \si{kg}^{-1})$ & $\Delta G / G (ppm)$ & Year & Ref. \\ \hline
1 & H. Cavendish & 316 & Lead & Sphere & 6.754 & 6000 & 1798 & ~[\onlinecite{cav1798}] \\
2 & F. Reich & 90 & Lead & Spheres & 6.64 & 4283 & 1838 & ~[\onlinecite{Reich1838}]\\
3 & P. Von Jolly & 5775 & Lead & Sphere & 6.447 & 17000 & 1881 & ~[\onlinecite{Jolly1881}] \\
4 & J. Wilsing & 650 & Cast iron & Cylinders & 6.594 & 2275 & 1889 & ~[\onlinecite{Wilsing1889}] \\
5 & J. H. Poynting & 150 & Lead & Sphere & 6.698 & 5970 & 1891 & ~[\onlinecite{Poynting1891}]\\
6 & C. V. Boys & 14.8 & Lead & Spheres & 6.658 & 1051 & 1895 & ~[\onlinecite{Boys1895}] \\
7 & R. Eötvös & 600 & Lead & Rect. block & 6.657 & 1953 & 1896 & ~[\onlinecite{Roland1896}] \\
8 & C. Braun & 18.0 & Mercury & Spheres & 6.658 & 300 & 1897 & ~[\onlinecite{Braun1897}] \\
9 & Richarz et al. & 100000 & Lead & Rect. block & 6.683 & 645 & 1897 & ~[\onlinecite{Richarz1898}] \\
10 & G. K. Burgess & 20 & Lead & Spheres & 6.64 & 6024 & 1899 & ~[\onlinecite{Burgess1899}]  \\
11 & J. Zahradn\'{i}\v{c}ek & 23 & Lead & Spheres & 6.659 & 6006 & 1933 & ~[\onlinecite{Zahradnicek1933}] \\
12 & Heyl et al. & 132 & Tool steel & Cylinders & 6.673 & 615 & 1942 & ~[\onlinecite{Heyl1942}] \\
13 & A. H. Cook & 1000 &  CuAl alloy & Cyl.assy & -- & -- & 1968 & ~[\onlinecite{Cook1968}] \\
14 & Rose et al. & 10.49 & Tungsten & Spheres & 6.674 & 1798 & 1969 & ~[\onlinecite{Rose1969}] \\
15 & Y. Renner & 35 & Mercury & Cylinders & 6.670 & 1199 & 1970 & ~[\onlinecite{Renner1970}] \\
16 & C. Pontikis & 3.0 & Ag, Cu, Pb, Hg & Spheres & 6.6714 & 90 & 1972 & ~[\onlinecite{Pontikis1972}] \\
17 & W. Koldewyn A. & 97 & Bronze & Axial doughnut & 6.575 & 25 875 & 1976 & ~[\onlinecite{Koldewyn1976}] \\
18 & Sagitov et al. & 80 & Stainless steel & Cylinders & 6.6745 & 120 & 1979 & ~[\onlinecite{Sagitov1979}] \\
19 & Luther et al. & 21 & Tungsten & Spheres & 6.6726 & 75 & 1982 & ~[\onlinecite{Luther1982}] \\
20 & C. C. Speake & 9.2 & Brass & Cylinder & 6.65 & 34587 & 1983 & ~[\onlinecite{Speake1983}] \\
21 & Liu et al. & 8.7 & Brass & Cylinder  & 6.660 & 3605 & 1987 & ~[\onlinecite{Liu1987}] \\
22 & Dousse et al. & 20 & Lead & Spheres & 6.6722 & 764 & 1987 & ~[\onlinecite{Dousse1987}] \\
23 & Saulnier et al. & 6.1 & Uranium & Polygons  & 6.65 & 13 534 & 1989 & ~[\onlinecite{Saulnier1989}] \\
24 & Ritter et al. & 0.09 & DyFe & Cylinders & 6.67 & 23988 & 1990 & ~[\onlinecite{Ritter1990}] \\
25 & Yang et al. & 48224 & Water & Cyl. tank & 6.672 & 9967 & 1991 & ~[\onlinecite{Yang1991}] \\
26 & Michaelis et al. & 0.24 & Zerodur & Cylinders & 6.7174 & 298 & 1995 & ~[\onlinecite{Michaelis1995}] \\
27 & Michaelis et al. & 1.8 & Tungsten & Cylinders & 6.7154 & 83 & 1995 & ~[\onlinecite{Michaelis1995}] \\
28 & Luo et al. & 6.25 & Stainless steel & Cylinders & 6.6699 & 105 & 1998 & ~[\onlinecite{Luo1998}] \\
29 & Schwarz et al. & 521 & Tungsten alloy & Cyl. assy & 6.6873 & 1406 & 1999 & ~[\onlinecite{Schwarz1999}] \\
30 & Nolting et al. & 1000 & Water & Cyl. tank & 6.6754 & 220 & 1999 & ~[\onlinecite{Nolting1999}] \\
31 & Gundlach et al. & 33 & Stainless steel & Sph. assy & 6.674215 & 14 & 2000 & ~[\onlinecite{Gundlach2000}] \\
32 & Quinn et al. & 46 & Cu 0.7\% Te & Cylinders & 6.67559 & 41 & 2001 &  ~[\onlinecite{Quinn2001}] \\
33 & U. Kleinvo\ss & 1152 & Brass & Cylinders & 6.67422 & 150 & 2002 &  ~[\onlinecite{Kleinvoss2002}] \\
34 & Armstrong et al. & 54 & Cu and stainless steel & Cylinders & 6.67387 & 41 & 2003 &  ~[\onlinecite{Armstrong2003}] \\
35 & Baldi et al. & 281 & Stainless steel & Cylinder & 6.675 & 1048 & 2005 &  ~[\onlinecite{Baldi2005}] \\
36 & Hu et al. & 12.5 & Stainless steel & Cylinders & 6.6723 & 130 & 2005 &  ~[\onlinecite{Hu2005}] \\
37 & Schlamminger et al. & 13520 & Smercury  & Cyl. tank & 6.674252 & 18 & 2006 &  ~[\onlinecite{Schlamminger2006}] \\
38 & Fixler et al. & 540 & Lead & Axial doughnut & 6.693 & 5110 & 2007 &  ~[\onlinecite{Fixler2007}] \\
39 & Lamporesi et al. & 516 & Tungsten & Cylinders & 6.667 & 1710 & 2008 &  ~[\onlinecite{Lamporesi2008}] \\
40 & Luo et al. & 6.15 & Stainless steel & Spheres & 6.67349 & 27 & 2009 & ~[\onlinecite{Luo2009}] \\
41 & Tu et al. & 1.6  & Stainless steel & Spheres & 6.67349 & 26 & 2010 &  ~[\onlinecite{Tu2010}] \\
42 & Parks et al. & 480 & Tungsten & Cyl. assy & 6.67234 & 21 & 2010 &  ~[\onlinecite{Parks2010}] \\
43 & Quinn et al. & 45 & Cu 0.7\% Te & Cylinders & 6.67545 & 27 & 2013 &  ~[\onlinecite{Quinn2013}] \\
44 & Rosi et al. & 516 & Tungsten alloy & Cylinders & 6.67191 & 150 & 2014 &  ~[\onlinecite{Rosi2014}] \\
45 & Newman et al. & 59 & Copper & Rings & 6.67433 & 19 & 2014 &  ~[\onlinecite{Newman2014}] \\
46 & Li et al. & 0.778 & Stainless steel & Spheres & 6.674484 & 12 & 2018 &  ~[\onlinecite{Li2018}] \\
47 & Westphal et al. & $9.2\times 10^{-5}$ & Gold & Sphere & 6.04 & 9934 & 2021 &  ~[\onlinecite{Westphal2021}] \\
48 & Brack et al. & 3.88 & Tungsten & Beam & 6.82 & 16129 & 2022 &  ~[\onlinecite{Brack2023}] \\
\end{tabular}
\end{ruledtabular}
\end{table*}

\begin{table*}
\caption{\label{tab:tableOPTO}Overview of recent key experiments with cavity optomechanical resonators in chronological order. Several types of resonators and detection methods are used by different groups in the field and are measured at different temperatures $T$. The table shows the resonance frequency $\omega_m$, mechanical quality factor $Q_m$ and the effective mass $m$ of the resonator.}
\begin{ruledtabular}
\begin{tabular}{cllcccccc}
 & Author(s) & Resonator & $\omega_m /2\pi$ (MHz) & $Q_m$ & $T$ (K) & $m$ (kg) & Year & Ref. \\ \hline
1 & Metzger et al. & Mirror & $7.30 \times 10^{-3}$ & $2 \times 10^3$ & 295 & $8.6 \times 10^{-12}$ & 2004 & ~[\onlinecite{Metzger2004}] \\
2 & Arcizet et al. & Mirror & 0.815 & $10^4$ & 295 & $1.9 \times 10^{-7}$ & 2006 & ~[\onlinecite{Arcizet2006}] \\
3 & Kleckner et al.& Mirror & $1.25 \times 10^{-2}$ & $1.37 \times 10^3$ & 295 & $2.4 \times 10^{-11}$ & 2006 & ~[\onlinecite{Kleckner2006}] \\
4 & Gigan et al & Mirror & 0.278 & $9 \times 10^3$ & 295 & $9.0 \times 10^{-12}$ & 2006 & ~[\onlinecite{Gigan2006}] \\
5 & Arcizet et al. & Mirror & 0.814 & $10^4$ & 295 & $1.9 \times 10^{-6}$ & 2006 & ~[\onlinecite{2Arcizet2006}] \\
6 & Schliesser et al. & Toroidal Microresonator & 57.8 & $2.89 \times 10^3$ & 300 & $1.5\times 10^{-11}$ & 2006 & ~[\onlinecite{Schliesser2006}] \\
7 & Corbitt et al. & Mirror & $1.72 \times 10^{-4}$ & $3.2 \times 10^3$ & 295 & $1.0 \times 10^{-3}$ & 2007 & ~[\onlinecite{Corbitt2007}] \\
8 & Corbitt et al. & Mirror & $1.27 \times 10^{-5}$ & $1.995 \times 10^4$ & 295 & $1.0 \times 10^{-3}$ & 2007 & ~[\onlinecite{2Corbitt2007}] \\
9 & Favero et al & Mirror & 0.547 & $1.059 \times 10^3$ & 300 & $1.1 \times 10^{-14}$ & 2007 & ~[\onlinecite{Favero2007}] \\
10 & Caniard et al. & Mirror & 0.711 & $1.6 \times 10^4$ & 295 & $7.4 \times 10^{-4}$ & 2007 & ~[\onlinecite{Caniard2007}]\\
11 & Thompson et al. & Beam or membrane & 0.134 & $1.1 \times 10^6$ & 294 & $3.9 \times 10^{-11}$ & 2008 & ~[\onlinecite{Thompson2008}] \\
12 & Gröblacher et al. & Mirror & 0.557 & $2 \times 10^3$ & 35 & $4.0 \times 10^{-11}$ & 2008 & ~[\onlinecite{Groblacher2008}] \\
13 & Mow-Lowry et al. & Mirror & $8.48 \times 10^{-5}$ & $4.45 \times 10^4$ & 300 & $6.9 \times 10^{-4}$ & 2008 & ~[\onlinecite{MowLowry2008}]  \\
14 & Schliesser et al. & Toroidal Microresonator & 74.0 & $5.7 \times 10^4$ & 295 & $1.0 \times 10^{-11}$ & 2008 & ~[\onlinecite{Schliesser2008}] \\
15 & Regal et al. & Superconducting Circuit & 0.237 & $2.3 \times 10^3$ & 0.040 & $2.0 \times 10^{-15}$ & 2008 & ~[\onlinecite{Regal2008}]\\
16 & Liu et al. & Mirror & $1040$ & $1.8 \times 10^2$ & 295 & $2.0 \times 10^{-17}$ & 2008 & ~[\onlinecite{Liu2008}] \\
17 & Li et al. & Photonic Crystal & 8.87 & $1.85 \times 10^3$ & 295 & $1.3 \times 10^{-15}$ & 2008 & ~[\onlinecite{Li2008}]\\
18 & Teufel et al. & Beam or membrane & 1.53 & $3 \times 10^5$ & 0.050 & $6.2 \times 10^{-15}$ & 2008 & ~[\onlinecite{Teufel2008}] \\
19 & Gröblacher & Mirror & 0.945 & $3 \times 10^4$ & 5.3 & $4.3 \times 10^{-11}$ & 2009 & ~[\onlinecite{Groblacher2009}] \\
20 & Schliesser et al. & Toroidal Microresonator & 65.0 & $2 \times 10^3$ & 1.65 & $7.0 \times 10^{-11}$ & 2009 & ~[\onlinecite{Schliesser2009}] \\
21 & Park et al. & Toroidal Microresonator & 118.6 & $3.4 \times 10^3$ & 1.4 & $2.8 \times 10^{-11}$ & 2009 & ~[\onlinecite{Park2009}] \\
22 & Lin et al. & Toroidal Microresonator & 8.53 & $4.07 \times 10^3$ & 300 & $1.5 \times 10^{-13}$ & 2009 & ~[\onlinecite{Lin2009}] \\
23 & Lee et al. & Toroidal Microresonator & 6.272 & $5.45 \times 10^2$ & 300 & $3 \times 10^{-8}$ & 2010 & ~[\onlinecite{Lee2010}] \\
24 & Rocheleau et al. & Superconducting Circuit & 6.30 & $10^6$ & 0.020 & $2.1 \times 10^{-15}$ & 2010 & ~[\onlinecite{Rocheleau2010}] \\
25 & Anetsberger et al. & Toroidal Microresonator & 8.30 & $3 \times 10^4$ & 300 & $3.7 \times 10^{-15}$ & 2010 & ~[\onlinecite{Anetsberger2010}] \\
26 & Teufel et al. & Superconducting Circuit & 10.69 & $3.6 \times 10^5$ & 0.020 & $4.8 \times 10^{-14}$ & 2011 & ~[\onlinecite{Teufel2011}] \\
27 & Kuhn et al. & Beam or membrane & 3.2 & $5\times 10^6$ & 300 & $2.5 \times 10^{-8}$ & 2011 & ~[\onlinecite{Kuhn2011}] \\
28 & Zheng et al. & Photonic Crystal & 65 & $3.76 \times 10^2$ & 300 & $6.11 \times 10^{-15}$ & 2012 & ~[\onlinecite{Zheng2012}] \\
29 & Serra et al. & Mirror & 0.085 & $2.6 \times 10^6$ & 4.5 & $3 \times 10^{-7}$ & 2012 & ~[\onlinecite{Serra2012}] \\
30 & Karuza et al. & Mirror & 0.36 & $1.22 \times 10^5$ & 300 & $4.5 \times 10^{-11}$ & 2013 & ~[\onlinecite{Karuza2013}] \\
31 & Torres et al. & Beam or membrane & 0.40 & $7.5 \times 10^5$ & 300 & $2.1 \times 10^{-7}$ & 2013 & ~[\onlinecite{Torres2013}] \\
32 & Doolin et al. & Mirror & 20.1 & $3.6 \times 10^3$ & 0.01 & $1.4 \times 10^{-16}$ & 2014 & ~[\onlinecite{Doolin2014}] \\
33 & Safavi-Naeini et al. & Photonic Crystal & $9.35\times 10^3$ & $3.74 \times 10^7$ & 20 & $4 \times 10^{-18}$ & 2014 & ~[\onlinecite{Naeini2014}] \\
34 & Song et al. & Beam or membrane & 24 & $1.5 \times 10^4$ & 0.06 & $5.9 \times 10^{-18}$ & 2014 & ~[\onlinecite{Song2014}] \\
35 & Paraïso et al. & Photonic Crystal & 11 & $10^7$ & 4 & $3.6 \times 10^{-15}$ & 2015 & ~[\onlinecite{Paraiso2015}] \\
36 & Pirkkalainen et al. & Superconducting Circuit & 13.032 & $3.9\times 10^4$ & 0.025 & $1.02 \times 10^{-13}$ & 2015 & ~[\onlinecite{Pirkkalainen2015}] \\
37 & Yuan et al. & Beam or membrane & 0.123 & $3.5\times 10^7$ & 0.000035 & $2 \times 10^{-10}$ & 2015 & ~[\onlinecite{Yuan2015}] \\
38 & Pontin et al. & Mirror & 0.172 & $5\times 10^4$ & 300 & $2.5 \times 10^{-7}$ & 2016 & ~[\onlinecite{Pontin2016}] \\
39 & Santos et al. & Beam or membrane & 7 & $10^7$ & 0.03 & $1.0 \times 10^{-7}$ & 2017 & ~[\onlinecite{Santos2017}] \\
40 & Cripe et al. & Mirror & $2.88\times 10^{-4}$ & $8\times 10^3$ & -- & $5 \times 10^{-10}$ & 2018 & ~[\onlinecite{Cripe2018}] \\
41 & Ockeloen-Korppi et al. & Superconducting Circuit & 10 & $10^5$ & 0.014 & $4.2 \times 10^{-14}$ & 2018 & ~[\onlinecite{Korppi2018}] \\
42 & Tavernarakis et al. & Beam or membrane & $0.038$ & 2245 & 300 & $7.9 \times 10^{-19}$ & 2018 & ~[\onlinecite{Tavernarakis2018}] \\
43 & Hauer et al. & Beam or membrane & 11.2 & $2.99\times 10^4$ & 4.2 & $1.83 \times 10^{-16}$ & 2019 & ~[\onlinecite{Hauer2019}] \\
44 & Rodrigues et al. & Beam or membrane & 7.129 & $9\times 10^5$ & 0.015 & $1.0 \times 10^{-15}$ & 2019 & ~[\onlinecite{Rodrigues2019}] \\
45 & Delić et al. & Levitated Mircosphere & 0.305 & $1.45\times 10^2$ & 0.000012 & $3.57 \times 10^{-18}$ & 2020 & ~[\onlinecite{Delić2020}] \\
46 & Lépinay et al. & Beam or membrane & 6.69 & $1.2\times 10^5$ & 0.01 & $4.2 \times 10^{-14}$ & 2020 & ~[\onlinecite{Lépinay2020}] \\
47 & Bothner et al. & Beam or membrane &  1.4315 & $1.95\times 10^5$ & 0.015 & $6.8 \times 10^{-15}$ & 2020 & ~[\onlinecite{Bothner2020}] \\
48 & Liu et al. & Beam or membrane & $1.7\times 10^{-3}$ & $10^7$ & 0.02 & $1.26 \times 10^{-6}$ & 2021 & ~[\onlinecite{Liu2021}] \\
49 & Cattiaux et al. & Beam or membrane & 15.1 & $1.5\times 10^5$ & 0.0005 & $4.2 \times 10^{-14}$ & 2021 & ~[\onlinecite{Cattiaux2021}] \\
50 & Militaru et al. & Levitated Mircosphere & $0.073$ & $1.83\times 10^3$ & -- & $1.22 \times 10^{-18}$ & 2022 & ~[\onlinecite{Militaru2022}] \\
51 & Youssefi et al. & Superconducting Circuit  & 2.142 & $4.98\times 10^5$ & 0.015 & $4.2 \times 10^{-12}$ & 2022 & ~[\onlinecite{Youssefi2022}] \\
52 & Bothner et al. & Superconducting Circuit  & 5.32 & $4\times 10^5$ & 0.015 & $1.9 \times 10^{-15}$ & 2022 & ~[\onlinecite{Bothner2022}] \\
53 & Reigue et al. & Beam or membrane & $1.25\times 10^{-3}$ & $10^5$ & 0.02 & $1.23 \times 10^{-14}$ & 2023 & ~[\onlinecite{Reigue2023}] \\
54 & Piotrowski et al. & Levitated Mircosphere & 0.23 & $2.3\times 10^3$ & 300 & $3.4 \times 10^{-18}$ & 2023 & ~[\onlinecite{Piotrowski2023}] \\
55 & Tenbrake et al. & Beam or membrane & 2.1 &  20 & 4 & $2.4 \times 10^{-12}$ & 2024 & ~[\onlinecite{Tenbrake2024}] \\
56 & Agafonova et al. & Mirror & $1.8\times 10^{-5}$ & $8.6\times 10^4$ & 0.00024 & $1.0 \times 10^{-6}$ & 2025 & ~[\onlinecite{Sofia2025}] \\
\end{tabular}
\end{ruledtabular}
\end{table*}
\twocolumngrid
\clearpage

\section{Calculations of $t_{pG}$}
\subsection{Calculations of $A_1,A'_1,B,B',C,C'$}\label{sec:cABC}
\setcounter{figure}{0} 

By applying the unitary transformation $\hat{U}=e^{i\omega_l \hat{a}^{\dagger}\hat{a} t}$, the Hamiltonian transforms to
\begin{equation}
\begin{aligned}
        \hat{H}'_{tot}&=\frac{1}{2 M_1}\hat{p}_1^2 +\frac{1}{2}M_1 \omega_1^2 \hat{x}_1^2 - \hbar \Delta\hat{a}^{\dagger}\hat{a} \\
        &- \hbar G_{om}\hat{x}_1 \hat{a}^{\dagger}\hat{a} -\frac{G M_1 M_2}{d^3}[x_s \cos (\omega_s t + \phi_s) - \hat{x}_1]^2 \\
        &+ [i\hbar \sqrt{\eta_c \kappa}(\alpha_{l}e^{-i\phi_l}+\alpha_{p}e^{-i\omega t -i\phi_p})\hat{a}^{\dagger}+{\rm h.c.}],\\
\end{aligned}
\end{equation}
where $\Delta = \omega_l - \omega_c$ and $\omega = \omega_p - \omega_l$.
Substituting $\hat{H}'_\text{tot}$ into the Langevin-Heisenberg equations
\begin{equation}
\left\{
\begin{aligned}
\frac{{\rm d}\hat{x}_1}{{\rm d}t} =& \frac{\hat{p}_1}{M_1},\\
\frac{{\rm d}\hat{p}_1}{{\rm d}t}=&-M_1{\omega'_1}^2 \hat{x}_1 -\gamma_1 \hat{p}_1 + \hbar G_{om} \hat{a}^{\dagger}\hat{a}\\
&- \frac{2 G M_1 M_2 x_s}{d^3}\cos (\omega_s t + \phi_s) + \hat{F}_{in},\\
\frac{{\rm d}\hat{a}}{{\rm d}t} =& [i(\Delta + G_{om} \hat{x}_1)-\frac{\kappa}{2}]\hat{a} + \sqrt{\eta_c \kappa}\alpha_{l}e^{-i\phi_l} \\
&+ \sqrt{\eta_c \kappa}\alpha_{p}e^{-i\omega t -i\phi_p} + \hat{a}_{in},\\
\end{aligned}
\right.
\end{equation}
where $\hat{F}_{in}$ and $\hat{a}_{in}$ are external stochastical noise and $\omega'_1 = \sqrt{\omega_1^2 - 2 G M_1 M_2 /M_1 d^3}$. When considering the average response, by taking $\langle \hat{F}_{in} \rangle = 0$ and $\langle \hat{a}_{in} \rangle =0$ one obtains
\begin{equation}
\left\{
\begin{aligned}
\frac{{\rm d}^2\langle \hat{x}_1 \rangle}{{\rm d}t^2}=&-{\omega'_1}^2\langle\hat{x}_1 \rangle -\gamma_1  \frac{{\rm d}\langle\hat{x}_1 \rangle}{{\rm d} t}  + \frac{\hbar G_{om}}{M_1} \langle\hat{a}\rangle^2\\
&- \frac{2 G M_2 x_s}{d^3}\cos (\omega_s t + \phi_s),\\
\frac{{\rm d}\langle\hat{a}\rangle}{{\rm d}t} =& [i(\Delta + G_{om} \langle\hat{x}_1\rangle)-\frac{\kappa}{2}]\langle \hat{a}\rangle + \sqrt{\eta_c \kappa}\alpha_{l}e^{-i\phi_l}\\
&+ \sqrt{\eta_c \kappa}\alpha_{p}e^{-i\omega t -i\phi_p}.
\end{aligned}
\right.
\end{equation}
The above non-linear equations can be approximately solved using first-order perturbation methods. Treating $-2 G M_2 x_s \cos (\omega_s t + \phi_s)/d^3$ and $\sqrt{\eta_c \kappa}\alpha_{p}e^{-i\omega t -i\phi_p}$ as perturbation terms. Spliting $\langle \hat{x}_1 \rangle$ and $\langle \hat{a}\rangle$ into stationary and perturbation parts
\begin{equation}
    \begin{aligned}
        \langle \hat{x}_1 \rangle = \overline{x}_1 + \delta x_1,\quad \langle \hat{a}\rangle = \overline{a} + \delta a ,
    \end{aligned}
\end{equation}
where $\overline{x}_1$ and $\overline{a}$ are the stationary solution satisfied the unperturbed equations
\begin{equation}
\left\{
\begin{aligned}
0=&-{\omega'_1}^2 \overline{x}_1 + \frac{\hbar G_{om}}{M_1} \overline{a}^2, \\
0 =& (i\overline{\Delta}-\frac{\kappa}{2})\overline{a} + \sqrt{\eta_c \kappa}\alpha_{l}e^{-i\phi_l},\\
\end{aligned}
\right.
\end{equation}
with $\overline{\Delta}=\Delta + G_{om}\overline{x}_1$ while $\delta x_1$ and $\delta a$ are the perturbations satisfied the first-order perturbation equation
\begin{equation}
\left\{
\begin{aligned}
\frac{{\rm d}^2\delta x_1}{{\rm d}t^2}=&-{\omega'_1}^2\delta x_1 -\gamma_1  \frac{{\rm d}\delta x_1}{{\rm d} t}  + \frac{\hbar G_{om}}{M_1}(\overline{a}^{*}\delta a + \overline{a}^{*}\delta a^{*}) \\
&- \frac{2 G M_2 x_s}{d^3}\cos (\omega_s t + \phi_s),\\
\frac{{\rm d}\delta a}{{\rm d}t} =& (i\overline{\Delta} -\frac{\kappa}{2})\delta a + i G_{om} \overline{a} \delta x_1 + \sqrt{\eta_c \kappa}\alpha_{p}e^{-i\omega t -i\phi_p}.
\end{aligned}
\right.
\label{eq:deltax1_and_deltaalpha}  
\end{equation}
By applying the variable substitution $e^{i\arg \overline{a}}\delta a'=\delta a$, and express $\cos(\omega_s t + \phi_s)$ as $(e^{i(\omega_s t + \phi_s)}+e^{-i(\omega_s t + \phi_s)})/2$ the Eq. (\ref{eq:deltax1_and_deltaalpha}) becomes
\begin{equation}
\left\{
\begin{aligned}
\frac{{\rm d}^2\delta x_1}{{\rm d}t^2}=&-{\omega'_1}^2\delta x_1 -\gamma_1  \frac{{\rm d}\delta x_1}{{\rm d} t}  + \frac{\hbar G_{om}}{M_1}|\overline{a}|(\delta a' + \delta a'^{*}) \\
&+ \frac{1}{2}(G^- + G^+),\\
\frac{{\rm d}\delta a'}{{\rm d}t} =& (i\overline{\Delta} -\frac{\kappa}{2})\delta a' + i G_{om} |\overline{a}| \delta x_1 + P^-\\
\end{aligned}
\right.
\label{eq:G_and_P}  
\end{equation}
where
\begin{equation}
\begin{aligned}
G^\pm = -\frac{2 G M_2 x_s}{d^3}e^{\pm i(\omega_s t + \phi_s)},\ P^\pm = \sqrt{\eta_c \kappa}\alpha_{p}e^{\pm i(\omega t +\phi_p +\arg \overline{a})}.
\end{aligned}
\end{equation}
By taking the ansatz
\begin{equation}
\left\{
\begin{aligned}
\delta x_1 =& A_1 P^- + A'_1 G^- + A_1^* P^+ + {A_1'}^* G^+,\\
\delta a' =& B P^- + B' G^- + C P^+ + C' G^+,
\end{aligned}
\right.
\end{equation}
and substituting it into Eq. (\ref{eq:G_and_P}), one obtains
\begin{equation}
\left\{
\begin{aligned}
&-\omega^2 A_1 = -{\omega'_1}^2 A_1 +i\omega\gamma_1 A_1 + \frac{\hbar G_{om}}{M_1}|\overline{a}|(B + C^*), \\
&-i\omega B = (i\overline{\Delta} -\frac{\kappa}{2})B + i G_{om} |\overline{a}|A_1 + 1,\\
&-\omega^2A_1^* = -{\omega'_1}^2 A_1^* -i\omega \gamma_1 A_1^* + \frac{\hbar G_{om}}{M_1}|\overline{a}|(C + B^*) ,\\
&i\omega C = (i\overline{\Delta} -\frac{\kappa}{2})C + i G_{om} |\overline{a}|A_1^*,\\
&-\omega_s^2 A'_1 = -{\omega'_1}^2 A'_1  +i\omega_s \gamma_1 A'_1 + \frac{\hbar G_{om}}{M_1}|\overline{a}|(B' + C'^*) + \frac{1}{2},\\
&-i\omega_s B'= (i\overline{\Delta} -\frac{\kappa}{2})B' + i G_{om} |\overline{a}|A'_1 ,\\
&-\omega_s^2{A_1'}^*= -{\omega'_1}^2 {A_1'}^* -i\omega_s\gamma_1 {A_1'}^* + \frac{\hbar G_{om}}{M_1}|\overline{a}|(C' + B'^*)+ \frac{1}{2},\\
&i\omega_s C' = (i\overline{\Delta} -\frac{\kappa}{2})C' + i G_{om} |\overline{a}|{A_1'}^* ,\\
\end{aligned}
\right.
\end{equation}
which yields
\begin{equation}
\left\{
\begin{aligned}
A_1 =& \frac{\hbar G_{om} |\overline{a}|\chi (\omega)}{-i(\overline{\Delta}+\omega)+\frac{\kappa}{2}+2\overline{\Delta}f(\omega)},\\
B =& \frac{1+i f(\omega)}{-i(\overline{\Delta}+\omega)+\frac{\kappa}{2}+2\overline{\Delta}f(\omega)},\\
C^* =& \frac{-i \hbar G_{om}^2 |\overline{a}|^2 \chi (\omega)}{[i(\overline{\Delta}-\omega)+\frac{\kappa}{2}][-i(\overline{\Delta}+\omega)+\frac{\kappa}{2}+2\overline{\Delta}f(\omega)]},\\
A'_1 =& \frac{\frac{1}{2}[\frac{\kappa}{2}-i(\overline{\Delta}+\omega_s)]M_1 \chi (\omega_s)}{-i(\overline{\Delta}+\omega_s)+\frac{\kappa}{2}+2\overline{\Delta}f(\omega_s)},\\
B' =& \frac{\frac{i}{2} G_{om} |\overline{a}| M_1 \chi (\omega_s)}{-i(\overline{\Delta}+\omega_s)+\frac{\kappa}{2}+2\overline{\Delta}f(\omega_s)},\\
C'^* =& \frac{-\frac{i}{2} [-i(\overline{\Delta}+\omega_s)+\frac{\kappa}{2}]\hbar G_{om} |\overline{a}| M_1 \chi (\omega_s)}{[i(\overline{\Delta}-\omega_s)+\frac{\kappa}{2}][-i(\overline{\Delta}+\omega_s)+\frac{\kappa}{2}+2\overline{\Delta}f(\omega_s)]},\\
\end{aligned}
\right.
\end{equation}
where
\begin{equation}
\begin{aligned}
\chi (\omega) =& \frac{1}{M_1 (\omega_1^2 - \omega^2 - i\omega \gamma_1)},\\
f(\omega) =& \hbar G_{om}^2 |\overline{a}|^2 \frac{\chi (\omega)}{i(\overline{\Delta}-\omega)+\frac{\kappa}{2}}.
\end{aligned}
\end{equation}

\subsection{Calculations of $t_{pG}$ when $\omega_s = \omega$}
\label{sec:ctpG}
Since $\delta a$ depend on $B$ and $B'$ only at the frequency being considered, $C$ and $C'$ is neglected. Hence, $\delta a'$ can be expressed as
\begin{equation}
\begin{aligned}
\delta a' =& B P^- + B' G^-.
\end{aligned}
\end{equation}
In order to obtain a stationary spectrum, it is necessary to ensure that $\omega = \omega_s$. With the condition $\delta a$ becomes
\begin{equation}
\begin{aligned}
\delta a = \frac{1+i f(\omega)[1+re^{i\phi}\frac{2}{\kappa}(i\overline{\Delta}-i\omega + \frac{\kappa}{2})]}{-i(\overline{\Delta}+\omega)+\frac{\kappa}{2}+2\overline{\Delta}f(\omega)} \sqrt{\eta_c \kappa}\alpha_{p}e^{-i\omega t - i\phi_p},
\end{aligned}
\label{eq:expression for deltaalpha}
\end{equation}
where
\begin{equation}
    \begin{aligned}
        r = \frac{\kappa G M_1 M_2 x_s}{2 d^3 \hbar G_{om} |\overline{a}| \sqrt{\eta_c \kappa}\alpha_p},\ \phi = \arg \overline{a} + \phi_p - \phi_s -\pi.
    \end{aligned}
\end{equation}
The $t_{pG}$ is given by
\begin{equation}
    \begin{aligned}
        t_{pG} =& \frac{\alpha_{p}e^{-i\omega t - i\phi_p} - \sqrt{\eta_c \kappa}\delta a}{\alpha_{p}e^{-i\omega t - i\phi_p}}\\
        =& 1 - \frac{1+i f(\omega)[1+re^{i\phi}\frac{2}{\kappa}(i\overline{\Delta}-i\omega + \frac{\kappa}{2})]}{-i(\overline{\Delta}+\omega)+\frac{\kappa}{2}+2\overline{\Delta}f(\omega)}\eta_c \kappa.
    \end{aligned}
\end{equation}
\section{Calculations of $F_p$ an $F_G$}\label{sec:FpFG}
\setcounter{figure}{0} 

The variation of radiation pressure due to the perturbation of cavity field $\delta a$ as well as the equation satisfied by $\delta a$ are
\begin{equation}
    \begin{aligned}
        \delta F_{rad} = \hbar G_{om} (\overline{a}^* \delta a + \overline{a}\delta a^*),
    \end{aligned}
    \label{eq: expression for deltaF_rad}
\end{equation}
and
\begin{equation}
    \begin{aligned}
        \frac{{\rm d}\delta a}{{\rm d}t} =& (i\overline{\Delta} -\frac{\kappa}{2})\delta a + i G_{om} \overline{a} \delta x_1 + \sqrt{\eta_c \kappa}\alpha_{p}e^{-i\omega t -i\phi_p}.
    \end{aligned}
\end{equation}
Since $\delta a \propto e^{-i\omega t}$ as shown in Eq. (\ref{eq:expression for deltaalpha}), one obtains
\begin{equation}
    \begin{aligned}
    \delta a = \frac{i G_{om} \overline{a} \delta x_1 + \sqrt{\eta_c \kappa}\alpha_{p}e^{-i\omega t -i\phi_p}}{\frac{\kappa}{2} -i\overline{\Delta} - i\omega },
    \end{aligned}
    \label{eq:expression for deltaalpha in terms of deltax_1}
\end{equation}
Substituting Eq. (\ref{eq:expression for deltaalpha in terms of deltax_1}) into Eq. (\ref{eq: expression for deltaF_rad}) one obtains
\begin{equation}
    \begin{aligned}
         \delta F_{rad} = k_p\delta x_1 + F_p \cos (\omega t +\phi_{fp}),\\
    \end{aligned}
\end{equation}
where
\begin{equation}
    \begin{aligned}
         &k_p = i\hbar G_{om}^2 |\overline{a}|^2(\frac{1}{\frac{\kappa}{2} -i\overline{\Delta} - i\omega}-\frac{1}{\frac{\kappa}{2} +i\overline{\Delta} + i\omega}),\\
         &F_p \cos (\omega t +\phi_{fp}) = 2\Re (\frac{\hbar G_{om}\sqrt{\eta_c \kappa}\overline{a}^*\alpha_{p}}{\frac{\kappa}{2} -i\overline{\Delta} - i\omega}e^{-i\omega t -i\phi_p}).
    \end{aligned}
\end{equation}
The first term, proportional to $\delta x_1$, alters the natural frequency and damping of oscillator 1, which is known as the "optical spring" effect. The second term represents a periodic force with frequency $\omega$, $\propto \alpha_p$, which is the driving force caused by the probe tone. When $\omega = -\overline{\Delta}$, The amplitude and initial phase becomes $F_p=4\hbar G_{om}\sqrt{\eta_c \kappa}|\overline{a}|\alpha_{p}/\kappa$ and $\phi_{fp} = \phi_p +\arg \overline{a}$.
    
Similar, the variation of gravity on oscillator can be perturbatively expressed in terms of $\delta x_1$
\begin{equation}
    \begin{aligned}
         \delta F_{\rm grav} = k_G\delta x_1 + F_G \cos (\omega_s t +\phi_s + \pi),
    \end{aligned}
\end{equation}
where
\begin{equation}
    \begin{aligned}
         &k_G = \frac{2 G M_1 M_2}{d^3},\quad F_G = \frac{2 G M_1 M_2 x_s}{d^3}.
    \end{aligned}
\end{equation}
The first term, proportional to $\delta x_1$, is an antirestoring force causing a natural frequency shift $\omega_1 \rightarrow \omega'_1$ on oscillator 1 while the second term is the periodic gravitational driving force. The factor $\pi$ in the phase represents a phase difference of $\pi$ between the gravitational driving force and the displacement of the source mass $M_2$. This is due to the attractive nature of gravity, where the direction of the displacement of the source mass $M_2$ is opposite to the variation in the gravity it exerts on the test mass $M_1$ relative to the their equilibrium positions.
    
Based on the expression of $F_p$ and $F_G$ as well as $\phi_{fp}$ and $\phi_G$, one easily find that $r = F_G / F_p$ is the ratio of the amplitude of gravitational driving force to the amplitude of probe tone driving force at frequency $\omega = -\overline{\Delta}$ while $\phi = \phi_{fp} - \phi_G$ is the difference of the initial phase of gravitational driving force and the probe tone driving force.

\section{Expression for $|t_{pG}|^2$ and $|t_{p0}|^2$ when $\Delta' \rightarrow 0$}
\label{sec:tpGrw}
\begin{figure}[h]
    \centering    
    \hspace*{-0.1cm} 
    \vspace*{-0.5cm}
    \includegraphics[width=\linewidth]{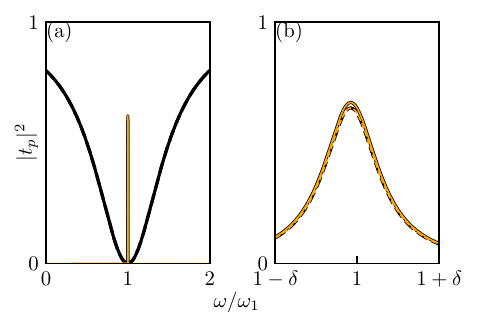}
    \caption{The figures shows the plot of $|t_p|^2$ corresponding to Eq.~\eqref{eq:tpGcondition}, ~\eqref{eq:tp0condition} and Eq.~\eqref{eq:tpGapprox}, ~\eqref{eq:tp0approx} with $\omega_1 = 2\pi\times 8\ \si{kHz}$ and $\kappa$, $g$, $Q_1$, $r$, $\phi$ at their reference values. The solid and dashed lines represent the curves for the presence and absence of gravitational driving, respectively. The curves in black correspond to Eq.~\eqref{eq:tpGcondition} and~\eqref{eq:tp0condition} without approximations while the curves in orange correspond to the approximation Eq.~\eqref{eq:tpGapprox} and~\eqref{eq:tp0approx}. In Fig.~\ref{fig:fig9}(a), the complete transmission spectra are presented while in Fig.~\ref{fig:fig9}(b), the spectra near $\omega = \omega'_1$ correspond to Fig.~\ref{fig:fig9}(a) are plotted with the $x$-direction magnified by a factor of $\delta = 5 \times 10^{-7}$ whose colors are consistent with those in Fig.~\ref{fig:fig9}(a).} 
    \label{fig:fig9}
\end{figure}

From Eq.~\eqref{eq:tpG} and~\eqref{eq:tp0}, when $\eta_c =1/2$ and $\overline{\Delta} = -\omega'_1$, the $t_{pG}$ and $t_{p0}$ can be expressed as
\begin{equation}
    \begin{aligned}
     t_{pG} = \frac{4g^2[(1+i\frac{\kappa}{4\omega'_1})(1+re^{i\phi})+\frac{\Delta'}{2\omega'_1} r e^{i\phi}] -2 i \Delta' y(\Delta')}{4g^2 + 2 (\frac{\kappa}{2} -i\Delta')y(\Delta')},
    \end{aligned}
    \label{eq:tpGcondition}
\end{equation}
and
\begin{equation}
    \begin{aligned}
     t_{p0} = \frac{4g^2 (1+i\frac{\kappa}{4\omega'_1}) -2 i \Delta' y(\Delta')}{4g^2 + 2 (\frac{\kappa}{2} -i\Delta')y(\Delta')},
    \end{aligned}
    \label{eq:tp0condition}
\end{equation}
where $\Delta' = \omega -\omega'_1$ and
\begin{equation}
    \begin{aligned}
     y(\Delta') =(1 + \frac{\Delta'}{2\omega'_1} + i\frac{\kappa}{4\omega'_1})[(1 + \frac{\Delta'}{\omega'_1})\gamma_1 - i(2 + \frac{\Delta'}{\omega'_1})\Delta'].
    \end{aligned}
\end{equation}
Since $\Delta'\rightarrow 0$, the term $-2i\Delta' y(\Delta')$ in the numerator and the term $2 (\frac{\kappa}{2} -i\Delta')y(\Delta')$ in the denominator can be expanded in powers of $\Delta'$ and retained up to the first-order term which gives 
\begin{equation}
    \begin{aligned}
     -2i\Delta' y(\Delta') &\approx -2i\gamma_1 (1 + i\frac{\kappa}{4\omega'_1})\Delta',
    \end{aligned}
\end{equation}
and
\begin{equation}
    \begin{aligned}
     2 (\frac{\kappa}{2} -i\Delta')y(\Delta') &\approx \kappa \gamma_1 (1+i\frac{\kappa}{4\omega'_1}) \\
     &+ \kappa [\frac{2\gamma_1}{\omega'_1} + \frac{\kappa}{2\omega'_1} -2i (1 + \frac{\gamma_1}{\kappa} - \frac{\kappa \gamma_1}{8{\omega'_1}^2})]\Delta'.
    \end{aligned}
\end{equation}
With this approximation, expressions of $t_{pG}$ and $t_{p0}$ become
\begin{equation}
    \begin{aligned}
        t_{pG} \approx \frac{4g^2 [(1+i\frac{\kappa}{4\omega'_1})(1+re^{i\phi} -\frac{i\gamma_1 \Delta'}{2g^2})+\frac{\Delta'}{2\omega'_1} r e^{i\phi}]}{z_1 + z_2\Delta'},
    \end{aligned}
        \label{eq:tpGapprox1}
\end{equation}
and
\begin{equation}
    \begin{aligned}
        t_{p0} \approx \frac{4g^2 (1+i\frac{\kappa}{4\omega'_1})}{z_1 + z_2\Delta'},
    \end{aligned}
        \label{eq:tp0approx1}
\end{equation}
where
\begin{equation}
    \begin{aligned}
    z_1 &=4g^2 + \kappa \gamma_1 + \frac{\kappa^2 \gamma_1}{4\omega'_1}i,\\
    z_2 &= \kappa (\frac{2\gamma_1}{\omega'_1} + \frac{\kappa}{2\omega'_1}) -2\kappa (1 + \frac{\gamma_1}{\kappa} - \frac{\kappa \gamma_1}{8{\omega'_1}^2})i.
    \end{aligned}
\end{equation}
Further, since $\Delta' \ll \omega'_1, \kappa, g$, $\Delta' \sim \gamma_1$ and $r < 1$ the terms $\Delta' r e^{i\phi}/2\omega'_1$ and $i\gamma_1 \Delta' /2g^2$ can be neglected while $z_1$ and $z_2$ can be approximate by
\begin{equation}
    \begin{aligned}
    z_1 &\sim 4g^2 + \kappa \gamma_1 + \frac{\kappa^2 \gamma_1}{4\omega'_1}i,\\
    z_2 &\sim \frac{\kappa^2}{2\omega'_1} -2i\kappa.
    \end{aligned}
\end{equation}
Under these approximation, $t_{pG}$ and $t_{p0}$  can be approximate by
\begin{equation}
    \begin{aligned}
        t_{pG} &\approx \frac{4g^2(1+i\frac{\kappa}{4\omega'_1})(1+re^{i\phi})}{4g^2 + \kappa \gamma_1 + \frac{\kappa^2}{2\omega'_1}\Delta' +i(\frac{\kappa^2 \gamma_1}{4\omega'_1} -2\kappa\Delta')},
    \end{aligned}
    \label{eq:tpGapprox}
\end{equation}
and 
\begin{equation}
    \begin{aligned}
        t_{p0} &\approx \frac{4g^2(1+i\frac{\kappa}{4\omega'_1})}{4g^2 + \kappa \gamma_1 + \frac{\kappa^2}{2\omega'_1}\Delta' +i(\frac{\kappa^2 \gamma_1}{4\omega'_1} -2\kappa\Delta')}.
    \end{aligned}
    \label{eq:tp0approx}
\end{equation}
while the $|t_{pG}|^2$ and $|t_{p0}|^2$ can be obtained by
\begin{equation}
    \begin{aligned}
        |t_{pG}|^2 & \approx \frac{16g^4 |1+re^{i\phi}|^2}{(\frac{4g^2}{1+\frac{\kappa^2}{16{\omega'_1}^2}} +\kappa \gamma_1)^2 +4\kappa^2 (\Delta' +\frac{g^2}{2\omega'_1 (1+\frac{\kappa^2}{16{\omega'_1}^2})})^2},\\
    \end{aligned}
    \label{eq:t2pGapprox}
\end{equation}
and
\begin{equation}
    \begin{aligned}
        |t_{p0}|^2 & \approx \frac{16g^4}{(\frac{4g^2}{1+\frac{\kappa^2}{16{\omega'_1}^2}} +\kappa \gamma_1)^2 +4\kappa^2 (\Delta' +\frac{g^2}{2\omega'_1 (1+\frac{\kappa^2}{16{\omega'_1}^2})})^2}.
    \end{aligned}
    \label{eq:t2p0approx}
\end{equation} 
As seen in Fig.~\ref{fig:fig9}, curves for $|t_p|^2$ corresponding to Eq.~\eqref{eq:tpGcondition}, ~\eqref{eq:tp0condition} and Eq.~\eqref{eq:tpGapprox},~\eqref{eq:tp0approx} with $\omega_1 = 2\pi\times 8\ \si{kHz}$ and $\kappa$, $g$, $Q_1$, $r$, $\phi$ at their reference values are plotted. In Fig.~\ref{fig:fig9}(a), the solid and dashed lines represent the transmission spectra in the presence and absence of gravitational driving, respectively. The curves in black correspond to Eq.~\eqref{eq:tpGcondition} and~\eqref{eq:tp0condition} without any approximations while the curves in orange correspond to the approximation Eq.~\eqref{eq:tpGapprox} and~\eqref{eq:tp0approx}. In Fig.~\ref{fig:fig9}(b), the curves for $|t_p|^2$ correspond to the two cases in Fig.~\ref{fig:fig9}(a) and are plotted using the same color scheme, with the $x$-axis direction magnified by a factor of $\delta = 5 \times 10^{-7}$. It can be observed that the curves with approximation in Eq.~\eqref{eq:tpGapprox} and~\eqref{eq:tp0approx} closely matching the curves correspond to Eq.~\eqref{eq:tpGcondition} and~\eqref{eq:tp0condition} without approximations within their transmission windows.
\nocite{*}
\clearpage

%
\end{document}